\documentclass[utf8]{FrontiersinHarvard} % for articles in journals using the Harvard Referencing Style (Author-Date), for Frontiers Reference Styles by Journal: https://zendesk.frontiersin.org/hc/en-us/articles/360017860337-Frontiers-Reference-Styles-by-Journal
\usepackage{url,hyperref,lineno,microtype,subcaption}
\usepackage[onehalfspacing]{setspace}

\def\keyFont{\fontsize{8}{11}\helveticabold }
\def\firstAuthorLast{P.~De La Torre Luque {et~al.}} %use et al only if is more than 1 author
\def\Authors{P.~De La Torre Luque\,$^{1}$, D.~Gaggero\,$^{4}$, D.~Grasso\,$^{2,3}$ and A.~Marinelli$^{5,6,7}$} %and C.~Evoli\,$^{7,8}$ O.~Fornieri\,$^{7,8}$, 

\def\Address{$^{1}$Stockholm University and The Oskar Klein Centre for Cosmoparticle Physics, Alba Nova, 10691 Stockholm, Sweden \\
$^{2}$INFN Sezione di Pisa, Polo Fibonacci, Largo B. Pontecorvo 3, 56127 Pisa, Italy \\
$^{3}$ Dipartimento di Fisica, Universit\`a di Pisa, Polo Fibonacci, Largo B. Pontecorvo 3  \\
$^{4}$ Instituto de F\'isica Corpuscular, Universidad de Valencia and CSIC, Edificio Institutos de Investigac\'ion, Calle Catedr\'atico Jos\'e Beltr\'an 2, 46980 Paterna, Spain  \\
$^{5}$  Dipartimento di Fisica ``Ettore Pancini'', Università degli studi di Napoli ``Federico II'', Complesso Univ. Monte S. Angelo, I-80126 Napoli, Italy\\
$^{6}$ INFN - Sezione di Napoli, Complesso Univ. Monte S. Angelo, I-80126 Napoli, Italy\\ 
$^{7}$ INAF, Osservatorio Astronomico di Capodimonte, Salita Moiariello 16, Napoli, 80131, Italy
}

\def\corrAuthor{Pedro De la Torre Luque, Daniele Gaggero, Dario Grasso and Antonio Marinelli}

\def\corrEmail{pedro.delatorreluque@fysik.su.se;   daniele.gaggero@ific.uv.es; dario.grasso@pi.infn.it; antonio.marinelli@na.infn.it}

\begin{document}
\onecolumn
\firstpage{1}

\title[Prospects for detection of a Galactic diffuse neutrino flux]{Prospects for detection of a Galactic diffuse neutrino flux} 

\author[\firstAuthorLast]{\Authors} %This field will be automatically populated
\address{} %This field will be automatically populated
\correspondance{} %This field will be automatically populated

\extraAuth{}% If there are more than 1 corresponding author, comment this line and uncomment the next one.
%\extraAuth{corresponding Author2 \\ Laboratory X2, Institute X2, Department X2, Organization X2, Street X2, City X2 , State XX2 (only USA, Canada and Australia), Zip Code2, X2 Country X2, email2@uni2.edu}

\maketitle

\begin{abstract}
%Recently, a new model of inhomogeneous transport of Galactic cosmic rays (CR) was reported. 
A Galactic cosmic-ray transport model featuring non-homogeneous transport has been developed over the latest years. This setup is aimed at reproducing $\gamma$-ray observations in different regions of the Galaxy (with particular focus on the progressive hardening of the hadronic spectrum in the inner Galaxy) and was shown to be compatible with the very-high-energy $\gamma$-ray diffuse emission recently detected up to PeV energies. 
In this work, we extend the results previously presented to test the reliability of that model throughout the whole sky. To this aim, we compare our predictions with detailed longitude and latitude profiles of the diffuse $\gamma$-ray emission measured by Fermi-LAT for different energies and compute the expected Galactic $\nu$ diffuse emission, comparing it with current limits from the ANTARES collaboration.
We emphasize that the possible detection of a Galactic $\nu$ component will allow us to break the degeneracy between our model and other scenarios featuring prominent contributions from unresolved sources and TeV halos. %Finally, we discuss possible drawbacks of our models and explain possible ways to solve them.
\tiny
 \keyFont{ \section{Keywords:} Galactic cosmic rays, Cosmic-ray transport, diffuse gamma rays, High energy gamma rays, diffuse neutrinos, Galactic center} 
\end{abstract}

\section{Introduction}

The Tibet AS$\gamma$ and LHAASO collaborations recently provided the first evidence of a diffuse $\gamma$-ray emission from the Galactic plane up to energies reaching the PeV~\citep{TibetASgamma:2021tpz, Zhao:2021dqj}. Since this emission is expected to be originated by the interaction of cosmic ray (CR) particles with the interstellar medium (ISM) and interstellar radiation fields (ISRFs), the Tibet AS$\gamma$ and LHAASO measurements offer a new valuable handle to study the origin and the propagation of Galactic CRs at energies never probed before ($\gg 100 \, \mathrm{TeV}$). 

The presence of a truly diffuse $\gamma$-ray at $\sim \mathrm{PeV}$ is likely due to $\sim \mathcal{O}(10) \, \mathrm{PeV}$ CRs injected by galactic PeV accelerators that were active in the past, the so-called \textit{PeVatrons}. The ability to explore the \textit{knee} region $(E_{\mathrm{CR}} \sim \mathrm{few} \, \mathrm{PeV's})$ of the CR spectrum is of outstanding importance for our understanding of CR physics. Indeed, if we assume the conventional scenario of Supernova Remnants (SNRs) as sources of the bulk of Galactic CRs, it is a theoretical challenge to even achieve particle acceleration at the level of $\sim \mathcal{O}(100) \, \mathrm{TeV}$~\citep{1983A&A...125..249L}. To overcome this problem, stellar clusters have recently come back as a viable explanation for such high-energy particle acceleration~\citep{1983SSRv...36..173C}, although it is not clear up to what extent in the locally observed CRs. 

Whether the \textit{knee} in the CR spectrum is due to a change in the CR acceleration mechanism or to a transport effect (see {\textit e.g.} \cite{Knee}) is also matter of debate. Moreover -- since due to spallation losses at those energies CR reaching the Earth must be originated within few kpc's -- it is not even clear if this feature is representative of the whole CR Galactic population or is shaped by local effects. The detection of $\gamma$-ray Galactic diffuse emissions above 100 TeV may offer a new valuable handle to clarify those conundrums.   

Moreover, since that emission is likely to be dominated by hadronic processes -- this due both to the larger abundance of CR nuclei respect to leptons and the increasing IC and synchrotron losses which prevent very energetic leptons from getting far from the acceleration region   
%decreasing Inverse Compton (IC) cross-section above the TeV 
-- a corresponding diffuse Galactic $\nu$ emission is also expected (see \textit{e.g.} \citealt{Berezinsky:1992wr,Evoli:2007iy} and the more recent \citealt{Ahlers:2015moa,Gaggero:2015xza}).
Noticeably, after almost ten years the birth of high energy $\nu$ astronomy \citep{IceCube:2014stg}, the experimental search of the Galactic $\nu$ diffuse emission has just started \citep{ANTARES:2017nlh,ANTARES:2018nyb} and a detection hint ($2 \sigma$) was recently reported by the IceCube collaboration \citep{IceCube:2019lzm}. Forthcoming dedicated analysis of IceCube and ANTARES data as well as those of KM3NeT \citep{KM3NET2016}, presently under advanced construction, should soon provide stronger evidences. 
Interestingly, a recent analysis of the IceCube public track-like events above 200 TeV -- performed externally to the collaboration -- claimed a $4.1\, \sigma$ detection of a neutrino diffuse emission along the Galactic Plane (GP).

Neutrinos will offer a valuable complementary probe of the CR population of the Galaxy. In fact, they are not subject to absorption -- which is significant for $\gamma$-rays above 100 TeV (see below) -- and can allow to single out the CR hadron component. This may be especially helpful to quantify and subtract the contribution of unresolved sources to the observed $\gamma$-ray diffuse emission which is expected to be dominated by leptonic sources (see \textit{e.g.} \citet{Casanova:2007cf,Vecchiotti:2021yzk}).

%for different locations in the sky, and, therefore, to better estimate the expected neutrino emission along the Galactic plane. 

%this finding open new perspectives for the understanding of the origin and propagation of Galactic CR. 

The interpretation of those measurements needs to compare the observed $\gamma$-ray and $\nu$ emissions among themselves and against detailed simulated templates of the diffuse $\gamma$-ray and $\nu$ diffuse emissions of the Galaxy. 
Those simulations require advanced numerical packages to model CR transport and interactions for given source and interstellar gas and radiation distributions as inferred from astronomical observations of proper tracers.
Well known examples of these packages are the {\tt GALPROP} code \citep{Strong_1998} which was extensively used by the Fermi-LAT collaboration or the more recently developed {\tt PICARD} code \citep{KISSMANN201437}.
In this work we use the {\tt DRAGON2} code \citep{Evoli2017jcap,Evoli2018jcap} -- to model CR transport -- in combination with the recently released {\tt HERMES} \citep{Dundovic:2021ryb} -- to produce simulated spectra and maps of the $\gamma$ and $\nu$ diffuse emissions. 

%Here we focus mainly on the latter item (some details about the gas distribution will be given in Sec.\ref{sec:gamma_model}, though).   
%Due to the challenges this imposes to current theoretical models of transport of charged particles in the Galaxy it is crucial to carefully study different scenarios of propagation, specially towards the centre of the Galaxy.
%The neutral pion decay component estimated by the
%conventional model should be accompanied by a $\nu$... This means that $\nu$ goes directly with the hadronic interactions but not sensitive to electromagnetic part...

%We notice that while in the last years a huge experimental effort allowed to strongly reduce the uncertainties on the CR transport parameters within a few kpc's horizon around the Solar System, yet very little is known about CR propagation beyond that distance. This requires to consider different scenarios which while reproducing local observable may however predict quite different CR spectra in more distant regions as the inner Galactic plane or the Galactic center.  

%With this regard, the study of weakly interacting messengers such as $\gamma$-rays and $\nu$s originated in the scattering of CRs with the InterStellar Medium (ISM) and the IS Radiation Field (ISRF) offers a valuable probe of CR transport in those distant regions.

{\tt DRAGON2} is built to model CR transport under very general conditions. In particular, it allows to account for a dependence of the diffusion coefficient on rigidity and position which was invoked in order to explain the hardening of the $\gamma$-diffuse emission above $10$~GeV observed by Fermi-LAT in the inner GP \citep{Gaggero:2014xla, Lipari:2018gzn, GuoModel} and motivated theoretically in \citet{Cerri2017jcap}.  %This performance will be exploited here to further investigate the predictions of that scenario.

%In particular we will 
%test against a large set of $\gamma$-ray data sets 
%focus on a scenario assuming in-homogeneous CR transport properties which was shown to account for the hardening of $\gamma$-ray diffuse emission observed by Fermi-LAT at low Galactic longitudes \citep{Gaggero:2015xza,Fermi-LAT:2016zaq} which provides, respect to conventional homogeneous models, a much better description of Fermi-LAT data.

%In this contribution, we show that these measurements seem to favour an inhomogeneous transport of cosmic rays throughout the Galaxy, specially motivated by the measurements of the Fermi-LAT detector. Moreover, we discuss the relevance of non-uniform cosmic-ray transport scenarios and the implications of these results for cosmic-ray physics and show that the energy spectra measured by Tibet AS$\gamma$, LHAASO, ARGO-YBJ and Fermi-LAT in several regions of the sky can be consistently described in terms of the emission arising by the Galactic cosmic-ray ``sea''. We also comment on the impact of other possible contributions, as the $\gamma$-ray emission from TeV halos or unresolved sources.

Recently, we presented a model of inhomogeneous transport of CRs able to explain the current diffuse $\gamma$-ray data from the GeV to the PeV range throughout several regions of the sky~\cite{Luque:2022buq}. In this work, we provide more details about that model, examine the predicted $\nu$ emission expected from such model and compare it with the model-independent limits posed by the ANTARES experiment.

\section{The {\Large $\gamma$} - optimized models}{\label{sec:gamma_model}}

\textbf{Motivations:} The transport of CR particles in the Galaxy is far from being well understood. Due to the complexity of the microphysics that describes the interactions of relativistic charged particles with the magneto-hydro-dynamic fluctuations, a fully satisfactory theoretical framework that describes how CRs are accelerated within astrophysical sources and subsequently propagate through the Galaxy is still far from being reached. 
Moreover, the measurements of CR spectra are limited to the vicinity of the Earth, preventing us from having information about the transport in other parts of the Galaxy. In addition, precision measurements of the CR spectra are limited to the energy region between a few hundreds of MeV to a few TeV, even though CRs have been detected up to energies of the order of $10^{11}$~GeV and Galactic CRs are thought to be produced up to, at least, $10^6$~GeV.

Conventional transport models are built to reproduce the local CR data and typically assume a very simplistic view of the Galaxy in which the propagation parameters characterizing the transport of CRs are isotropic and homogeneous.
%the magnetic halo, where CRs are confined, to be a cylinder mapped by the coordinates $(R,z)$ -- the Galactic center being at $(0,0)$ -- and that the propagation parameters does not depend on Galactocentric radius $R$ and it is isotropic. 
In particular, in these models the diffusion tensor is reduced to a spatially independent scalar function of the particle rigidity. Since the source spectrum (i.e. injection of CRs just depend on the sources accelerating them and not on where are the sources) is not expected to depend on the position, in this scenario the propagated CR spectrum -- which is a convolution of those two quantities --  is also expected to be spatially invariant. 
The normalization and rigidity dependence of the diffusion coefficient $D(\rho)$
% -- spatial isotropy is also assumed at least in an effective sense (see \citet{Cerri2017jcap}) --  
are generally fixed by reproducing the observations on secondary/primary ratios of CR species (mainly the boron-to-carbon ratio).  

In fact, this simplified approach has been shaken when several independent analysis of the Fermi-LAT data found evidences of a hardening of the spectrum of the $\gamma$-ray diffuse emission at low Galactic longitudes \citep{Gaggero:2014xla,Acero2016apjs,Yang2016prd}.  
Following \citet{Gaggero:2014xla,Gaggero:2015xza,Luque:2022buq}, in this work we adopt a scenario that assumes a spatially dependent diffusion coefficient which reproduces that feature and matches Fermi-LAT data better than conventional models. For this reason we call it $\gamma$-\textit{optimized}.

%In \citet{Gaggero:2014xla,Cerri2017jcap} the authors argue that the hardening of the $\gamma$-ray diffuse emission found by Fermi-LAT above 10 GeV \citep{Fermi-LAT:2012edv,Fermi-LAT:2016zaq} can be explained in terms of a radially dependent CR spectrum.
%Based on that interpretation, in \citet{Gaggero:2015xza} the authors predict a diffuse flux that is much larger than what is expected by conventional transport models at low Galactic longitudes and very high energies $E \gg 1$ TeV. They show that this also allows to consistently reproduce Fermi-LAT and Milagro \citep{Abdo:2008if} measurements in the inner Galaxy, as well as \citep{Gaggero:2017jts} the H.E.S.S. observed diffuse-emission in the Central Molecular Zone \citep{HESS:2016pst} (although the CR population in that small region may be affected by one or more PeVatrons close to the Galactic Center (GC)).

\textbf{CR spectra in the Galaxy from the $\gamma$-\textit{optimized} model:}
In \citet{Luque:2022buq}, we model the transport of CRs with the {\tt DRAGON2} numerical code~\citep{Evoli2017jcap,Evoli2018jcap}
within two transport scenarios: the $\gamma$-\textit{optimized} model, where the spectral index of the diffusion coefficient $\left( \delta(R)\right)$ depends on the galactocentric distance ($R$), according to the trend hinted by Fermi-LAT diffuse data, and the \textit{Base} model, where the diffusion coefficient is spatially constant and tuned on local CR data.
Specifically, we assume $\delta = 0.5$ throughout the Galaxy for the {\it Base} scenario while it is parametrized as
$$
\delta(R) = 0.04({\rm kpc^{-1}}) \cdot R({\rm kpc}) + 0.17 \quad ({\rm for}\; R < 8.5\; {\rm kpc})\, ,
$$
for the {\it optimized} one. The latter behaviour allows to reasonably reproduce the dependence of the CR proton spectral index inferred from Fermi-LAT data \citep{Gaggero:2014xla,Acero2016apjs,Yang2016prd} for $R > 2$ kpc (below this radius the increasing errors do not allow to trace it any further and we just extrapolate to $R = 0$ kpc the behaviour inferred at larger radii). 
In both scenarios the normalization of the diffusion coefficient is tuned to reproduce the boron-over-carbon ratio as shown in Fig. ~\ref{fig:BC} (the other main primary and secondary CR local spectra are reproduced as well).

\begin{figure}[ht!]
\begin{center}
\includegraphics[width=8.5cm]{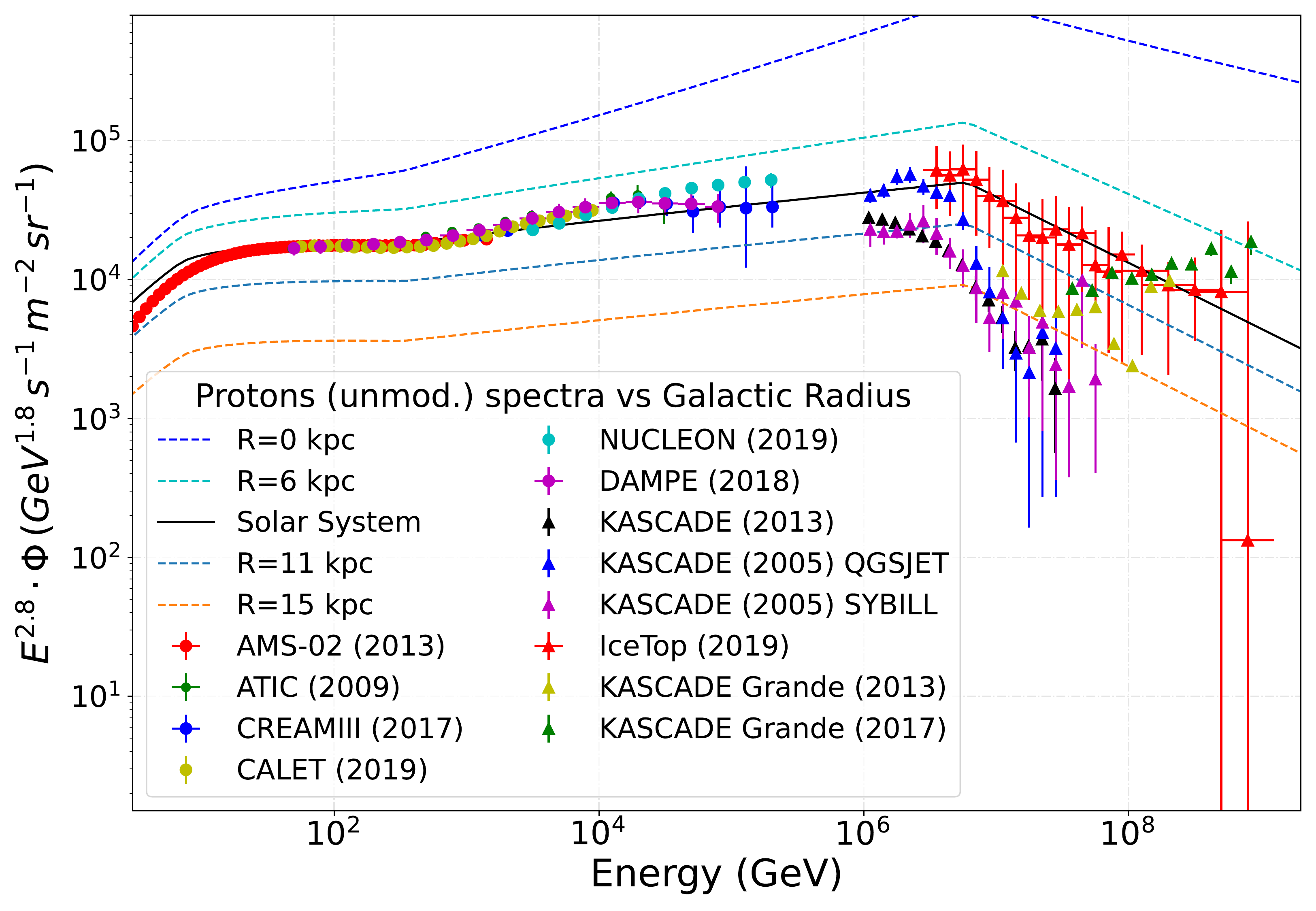}% 
\hspace{0.2cm}
\includegraphics[width=8.5cm]{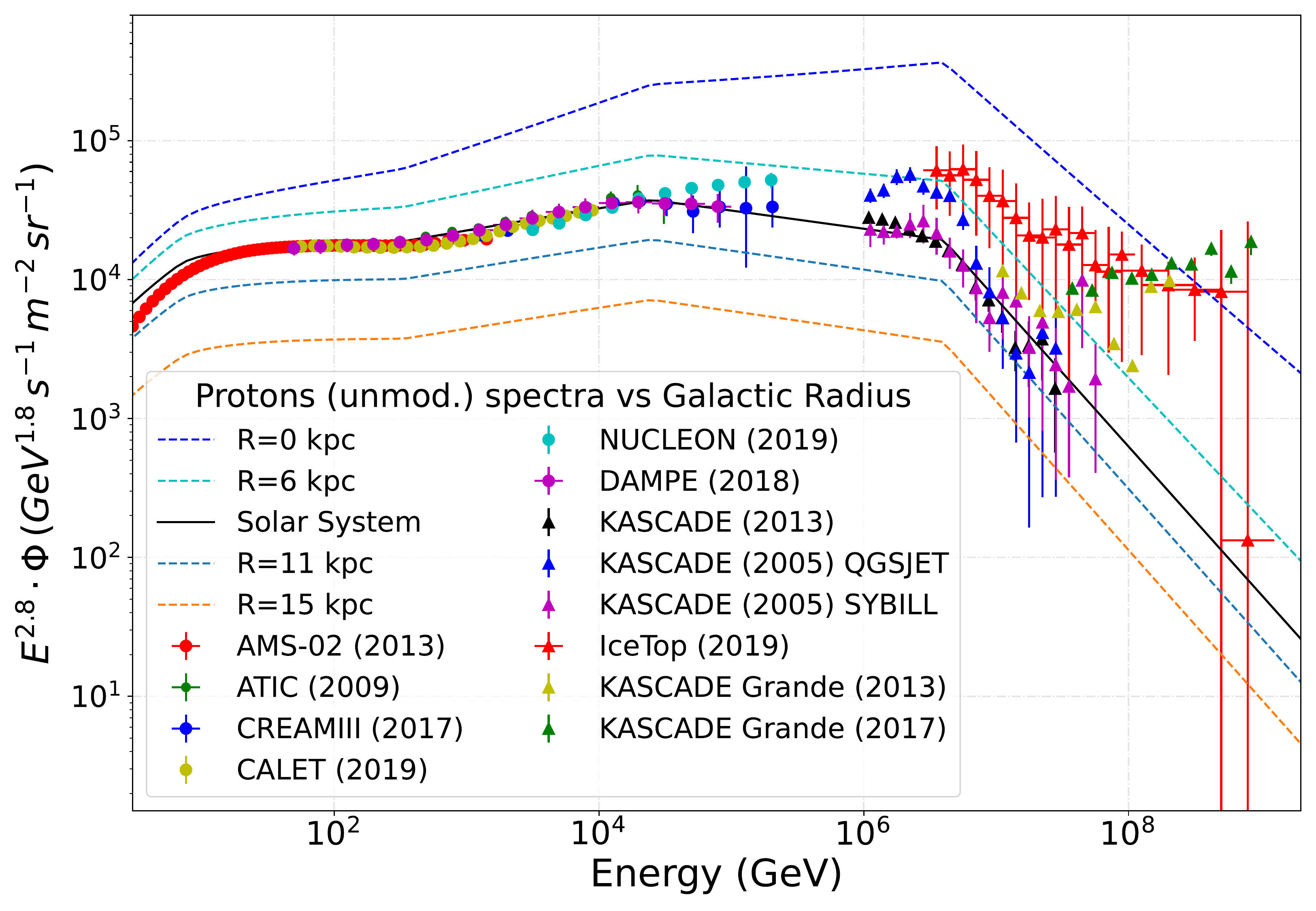}%   

\includegraphics[width=8.5cm]{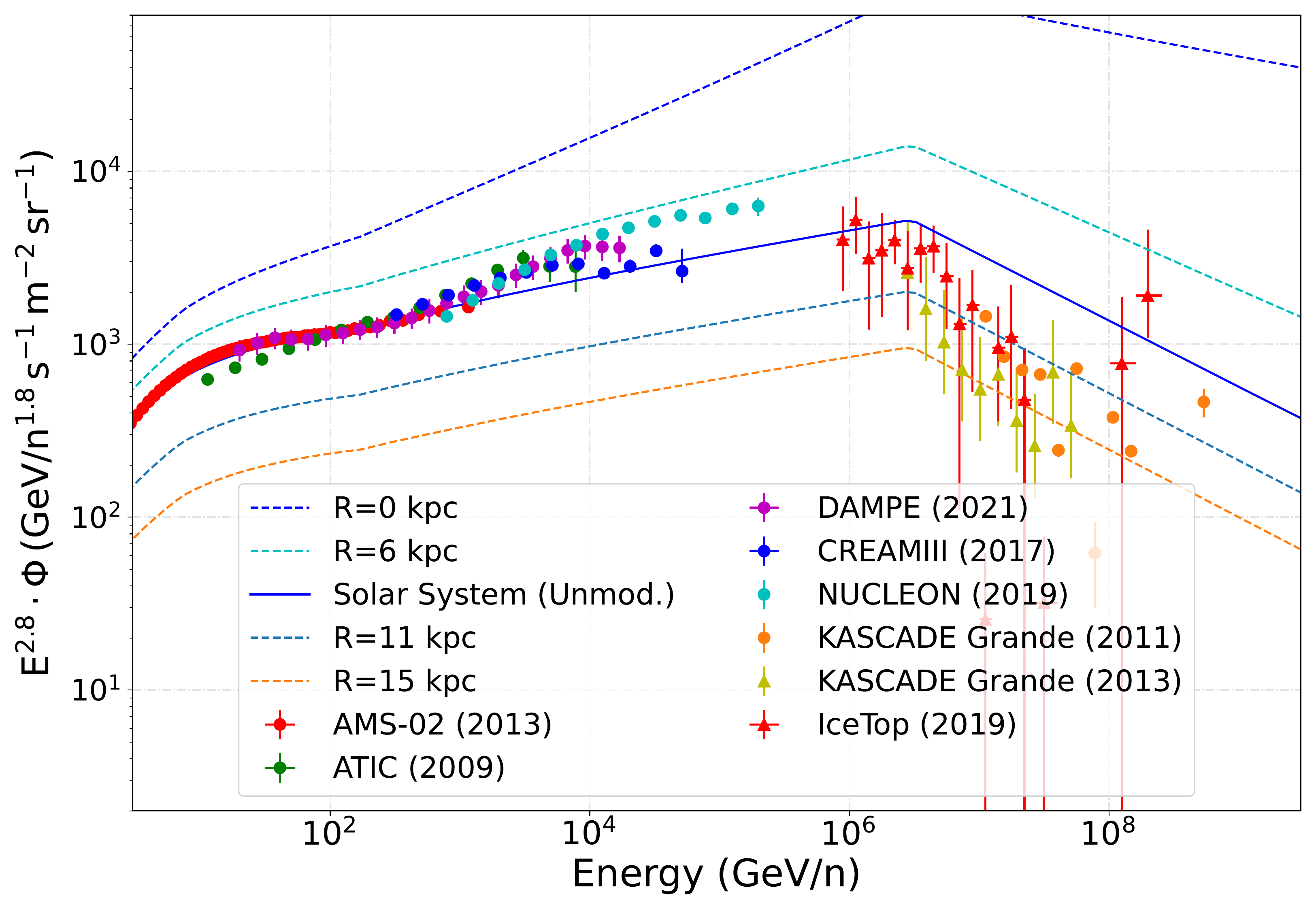}% 
\hspace{0.2cm}
\includegraphics[width=8.5cm]{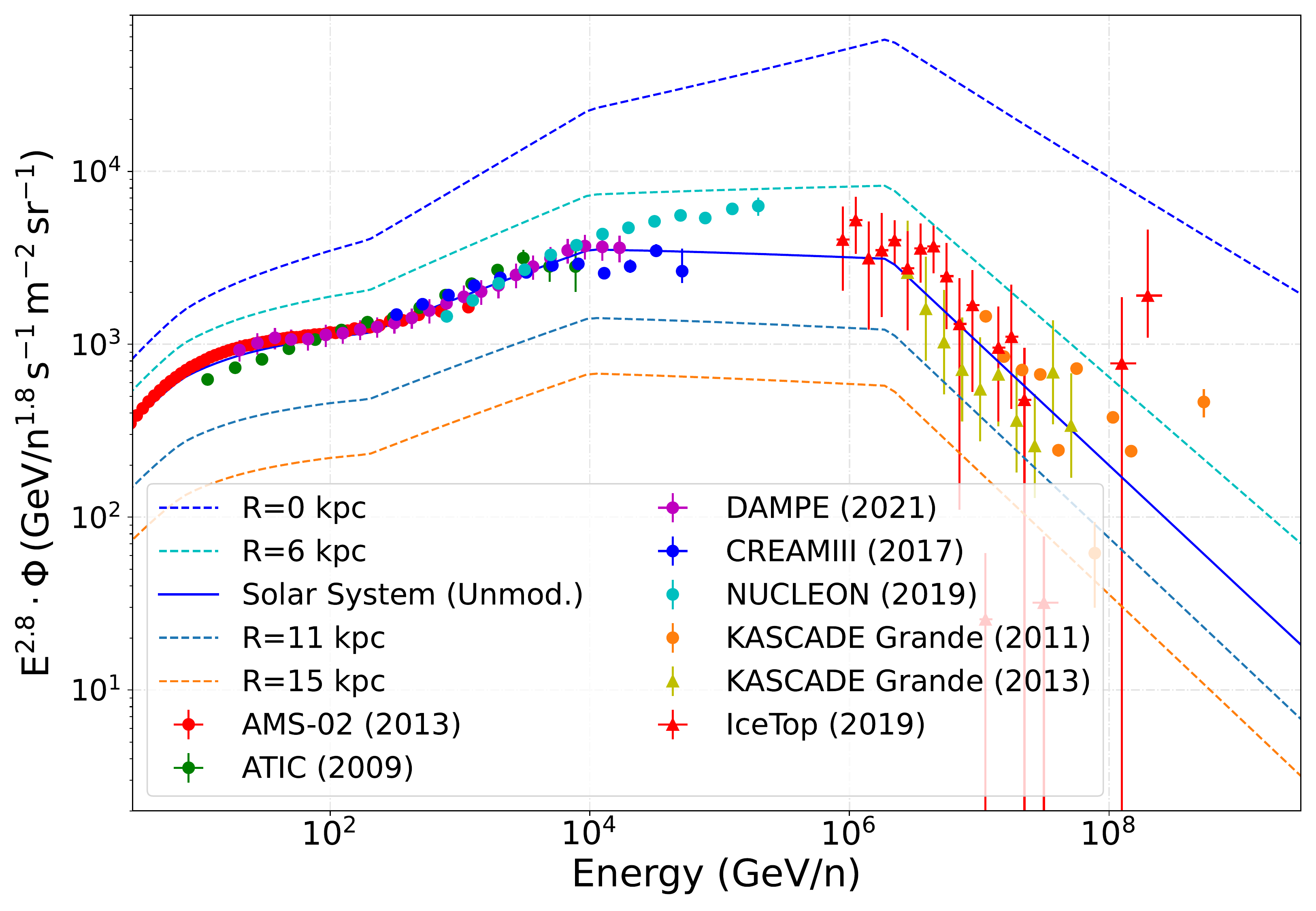}%   
\end{center}
\caption{Spectra of protons (upper panels) and helium (lower panels) of the $\gamma$-\textit{optimized} scenario for the Max (left panels) and Min (right panels) configurations, from $10$~GeV to $10^9$~GeV. 
 Since in the $\gamma$-\textit{optimized} scenario the propagation of CRs depends on the distance from the galactic center, we show the spectra at different galactocentric radii.
Available local CR data from AMS-02, ATIC, CREAM, CALET, NUCLEON, DAMPE, KASCADE, KASCADE Grande and IceTop are included for comparison.}
\label{fig:CRs}
\end{figure}

\begin{figure}[ht!]
\begin{center}
\includegraphics[width=8.5cm, height=6cm]{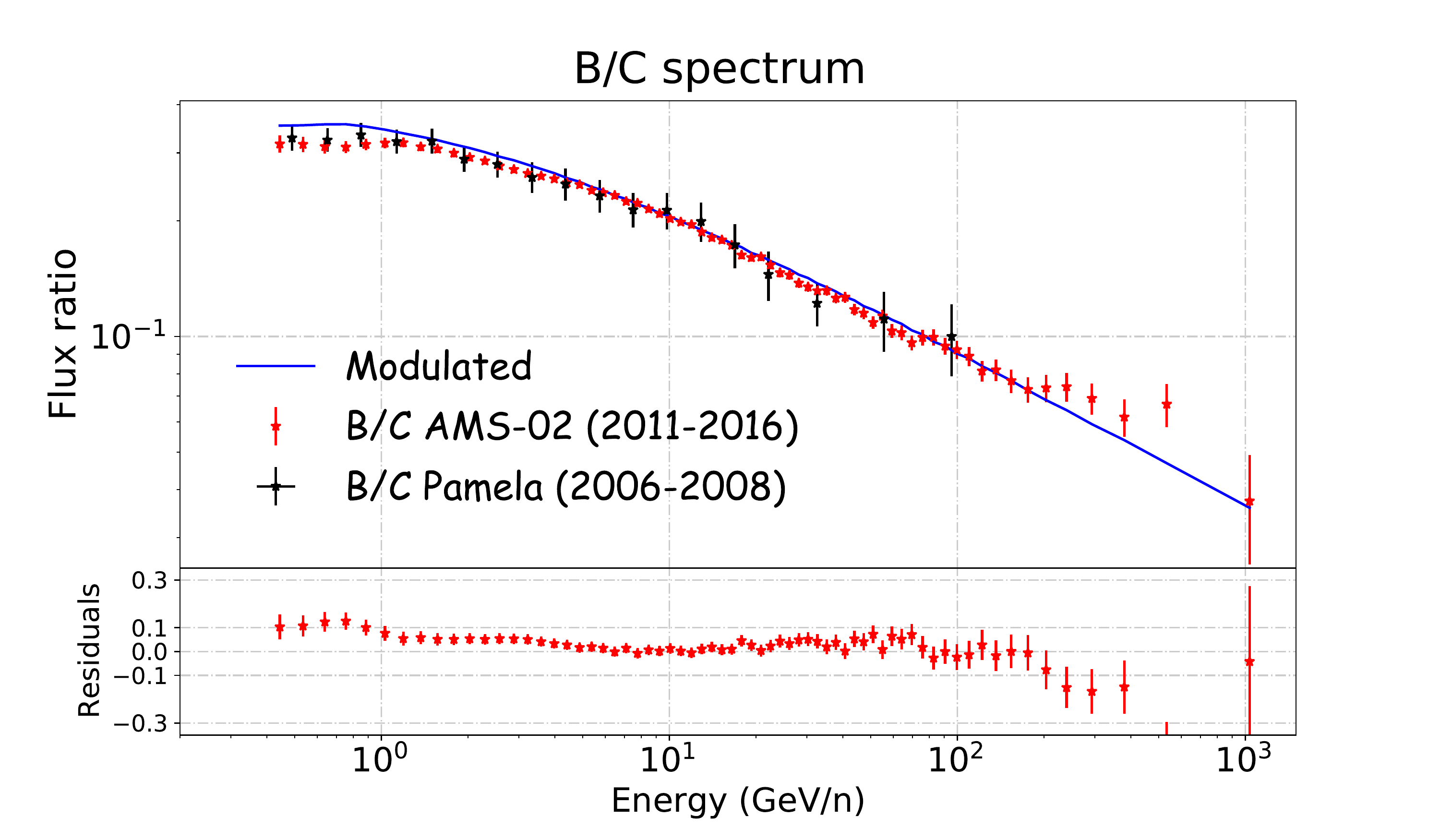}
\end{center}
\caption{ Boron-over-carbon spectrum predicted by the $\gamma$-\textit{optimized} model at Earth position ($\sim 8.3$~kpc from the galactic centre), compared to experimental data from Pamela and AMS-02.}
\label{fig:BC}
\end{figure}

Moreover, in order to extend our predictions to the highest energies, we account for a wide set of CR data in the PeV domain. In this context, we emphasize the large discrepancies in the energy spectra observed by different collaborations at these energies (see Fig.~\ref{fig:CRs}). In fact, these measurements suffer from large systematic errors, mostly associated with modelling hadronic interaction within the Earth atmosphere. Therefore, we consider two set-ups for the CR injection spectra (broken power-laws with spectral indexes and breaks reported in Table ~\ref{tab:diff_params}), which we call \textit{Min} and \textit{Max} configurations (see \citet{delaTorreLuque:2022vhm} for more details).

\begin{table*}[t]
    \centering
    \begin{tabular}{|c|c|c|c|c|c|c|c|c|}
    \hline
    \multicolumn{9}{c}{\textbf{Injection parameters}} \\
    \hline 
    \hline
    & $^1\mathbf{H} \; \gamma_1$ & $^1\mathbf{H} \; \gamma_2$ & $^1\mathbf{H} \; \gamma_3$ & $^1\mathbf{H} \; \gamma_4$ & $^4\mathbf{He} \; \gamma_1$ & $^4\mathbf{He} \; \gamma_2$ & $^4\mathbf{He} \; \gamma_3$ & $^4\mathbf{He} \; \gamma_4$ \\
    \hline
    Max model & 2.33 & 2.23 & 2.78 & --- & 3.28 & 2.18 & 2.69 & --- \\
    \hline
    Min model & 2.33 & 2.16 & 2.44 & 3.37 & 2.30 & 2.06 & 2.34 & 3.01 \\
    \hline
  \end{tabular}
  
  \caption{\small{Spectral indexes at injection for the Max and Min models. These spectral indexes are tuned to CR local data as described above and correspond to spectral breaks at the following energies: $335$ and $6 \cdot 10^6$~GeV for the Max models and $335$, $2\cdot 10^4$ and $4 \cdot 10^6$~GeV for the Min models. }}
    \label{tab:diff_params}
\end{table*}

Figure~\ref{fig:CRs} shows the spectra of protons (top panels) and Helium (bottom panels) on the Galactic plane at different distances from the centre for the $\gamma$-\textit{optimized} scenario. As a consequence of the radially-dependent diffusion coefficient adopted in that scenario -- we assume here the source spectra to be the same throughout the Galaxy -- the propagated spectra are significantly harder towards the centre of the Galaxy, while for the \textit{Base} scenario they have the same shape in every position although the normalization would vary depending on the density of sources at different regions of the Galaxy. In Figure ~\ref{fig:CRs} , the left panels show our predictions for the Max injection spectra setup while the right panels show the predicted spectra for the Min one.

In addition, in Figure~\ref{fig:BC} we show the boron-over-carbon spectrum obtained from the $\gamma$-\textit{optimized} model at Earth position, compared to the existent data. This observable is directly related to the details of the propagation of CRs and the ``grammage'' associated to the production of secondary CRs~\cite{delaTorreLuque:2022vhm, Luque:2021ddh, Luque:2021nxb}.
The combination of this piece of information with the $\gamma$-ray diffuse emission in the Galaxy may allow us to test whether features in the CR spectra --- such as the hardening at $\sim 300\,{\rm GeV/n}$ found by CREAM~\citep{Ahn_2010}, PAMELA~\citep{doi:10.1126/science.1199172}, AMS~\citep{PhysRevLett.114.171103} and the softening at $\sim 10\,{\rm TeV/n}$ measured by DAMPE~\citep{DAMPEBUMP} and CREAM~\citep{Yoon_2017} ---
are due to the injection or if they are representative of the whole Galaxy, namely due to transport effects~\citep{Blasi2012}.

%Moreover, the detection of the diffuse emission at low Galactic longitudes gives crucial information on how CR propagation behaves in the inner regions of the Galaxy, where the conditions for CR transport are expected to be different than in the average disc, noticeably leading to a radial (Galactocentric) dependence for the CR spectra.

\textbf{$\gamma$-ray profiles in the Galaxy from the $\gamma$-\textit{optimized} model:}
We use the recently released {\tt HERMES} code~\citep{Dundovic:2021ryb} to convolve along the line of sight the CR spatial and energy distributions modeled with {\tt DRAGON2}, updated gas (for the hadron emission) and ISRF (for the IC emission) models and the proper $\gamma$-ray cross-sections to get detailed full-sky maps of the expected diffuse emissions for each channel. 

In our previous work \citep{Luque:2022buq}, we tuned the CR propagation parameters to reproduce local CR data as well as  the diffuse emission measured by {\textit Fermi}-LAT for several GP quadrants. Then we used the same models to predict the emission at larger energies which we compared with ARGO-YBJ~\citep{ARGO-YBJ:2015cpa}, Tibet AS$\gamma$ data as well as with LHAASO preliminary results. 
%and with IceCube~\citep{IceCube:2019scr} and CASA-MIA~\citep{Borione:1997fy} upper limits between 100 TeV and 2 PeV in regions closer to the GC. 
We showed that those data in combination with {\textit Fermi}-LAT favor a spatially dependent transport scenario. 

\begin{figure}[ht!]
\begin{center}
\includegraphics[width=8.2cm]{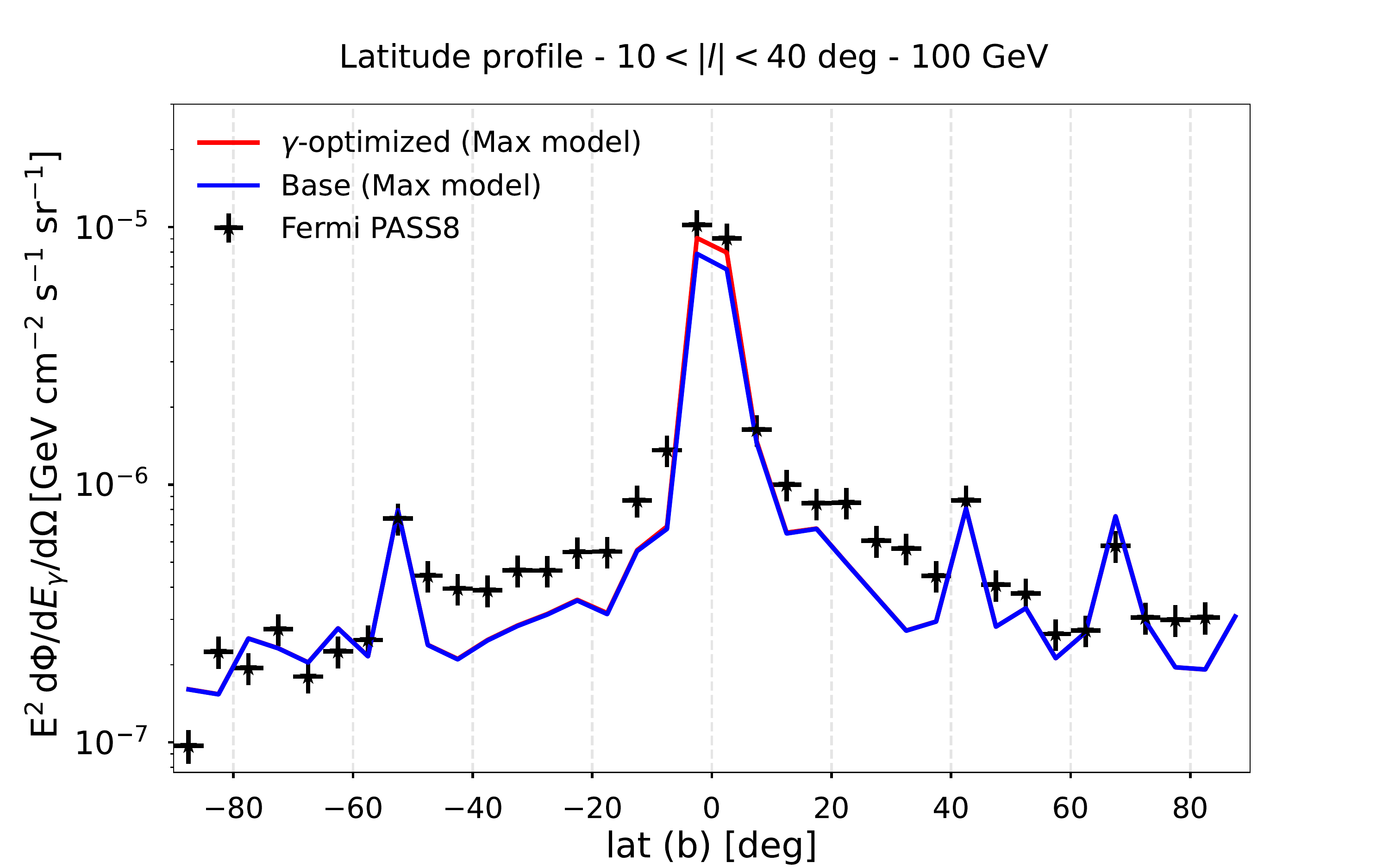}
\includegraphics[width=8.2cm]{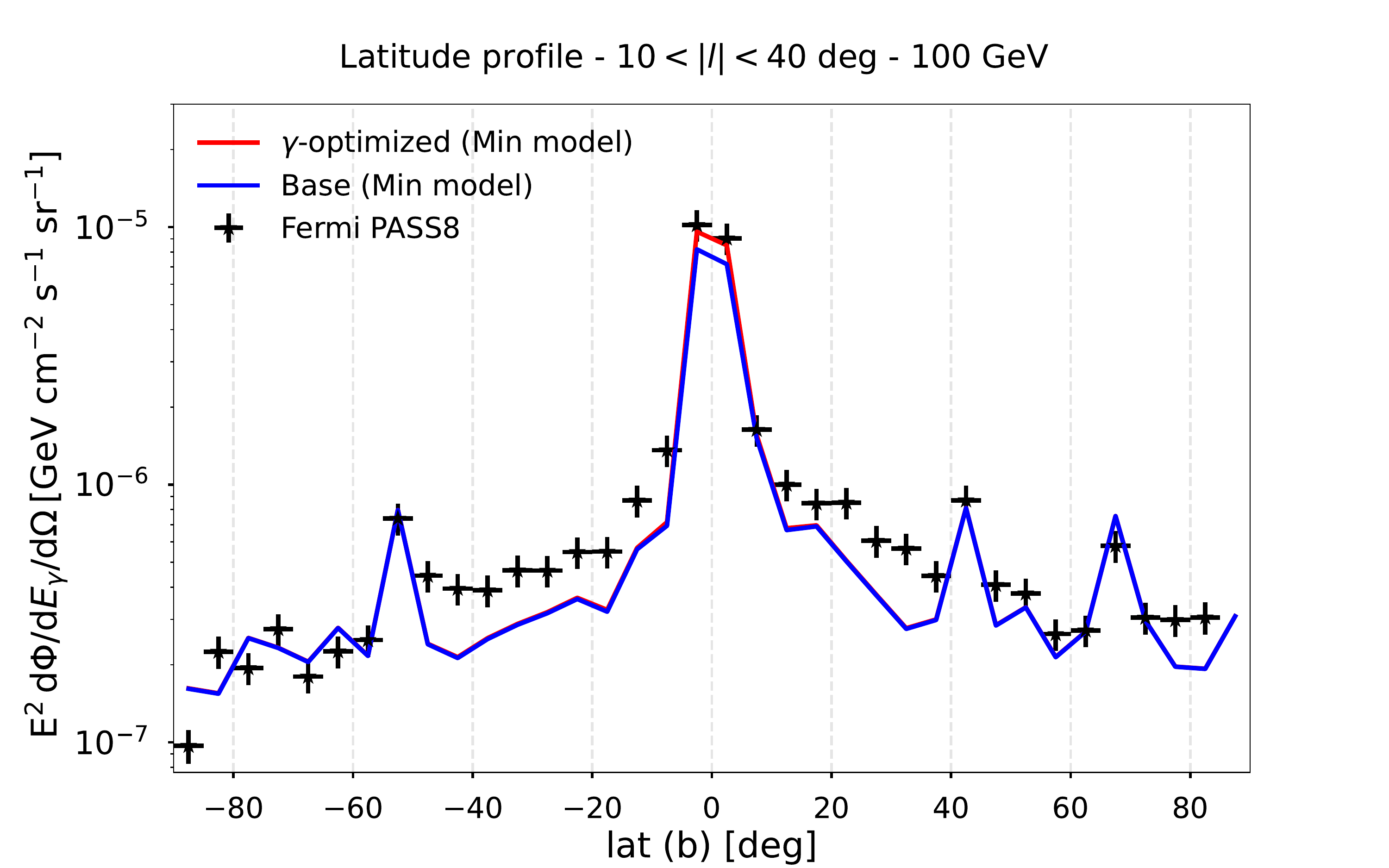}

\includegraphics[width=8.2cm]{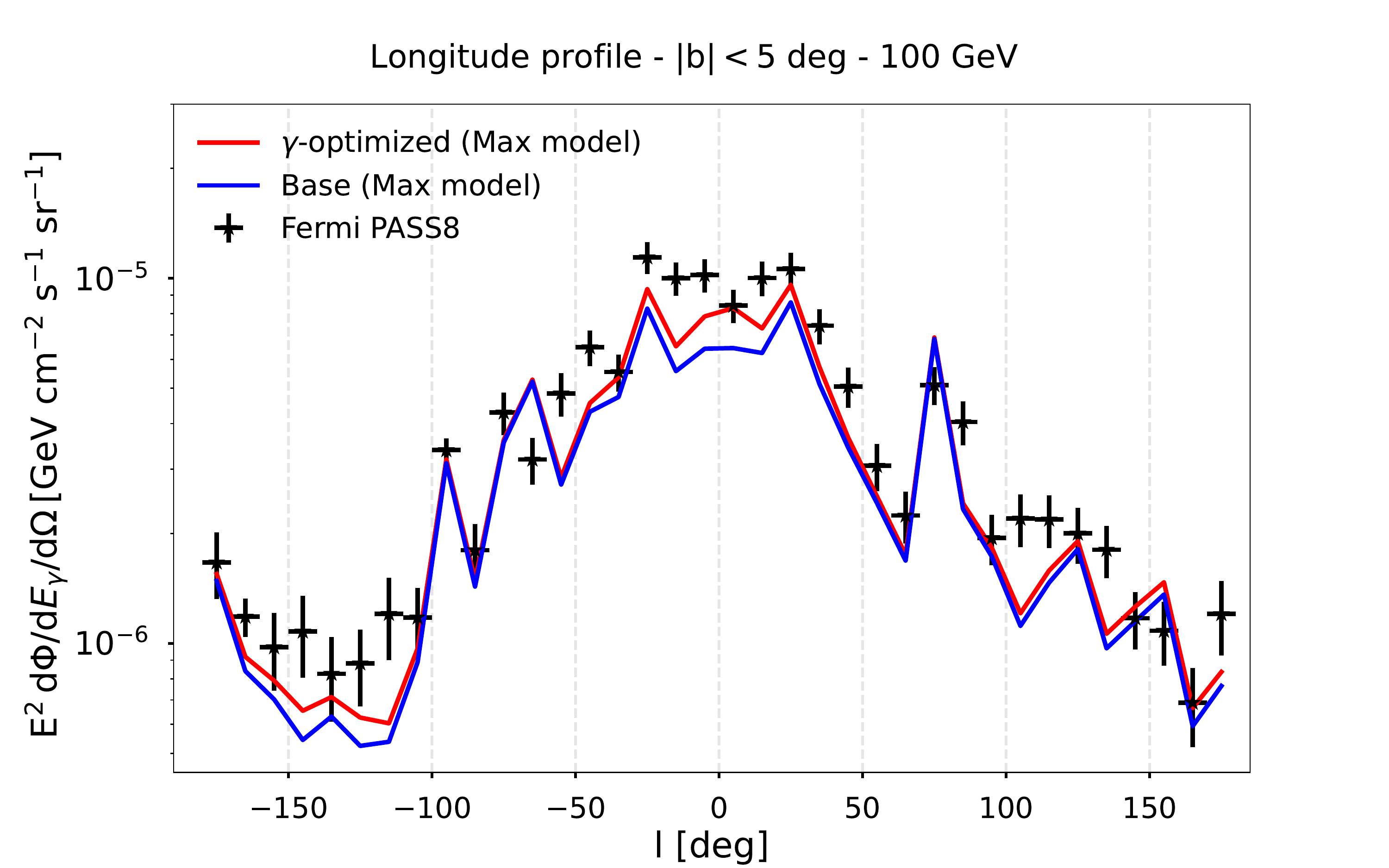}
\includegraphics[width=8.2cm]{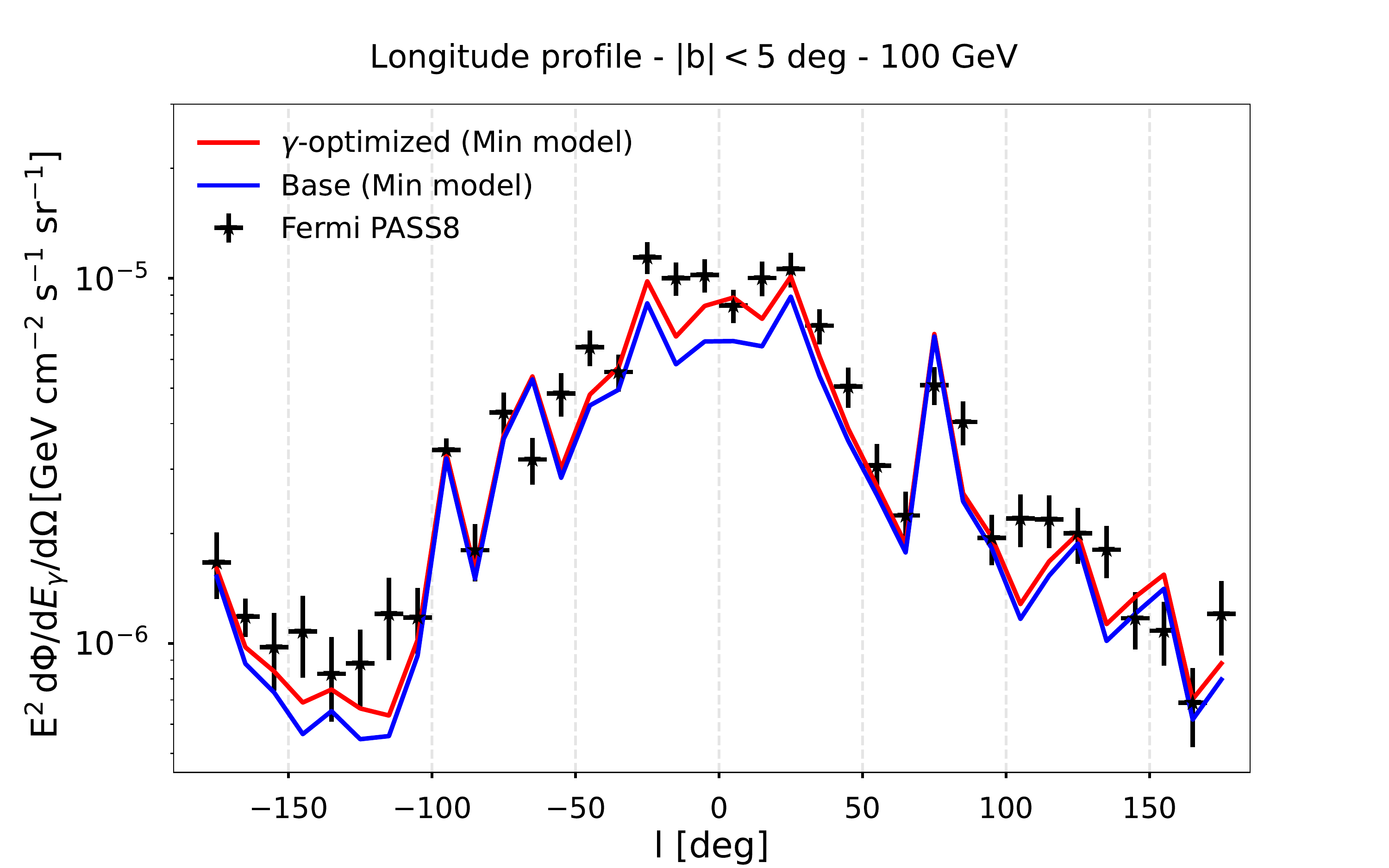}

\includegraphics[width=8.7cm]{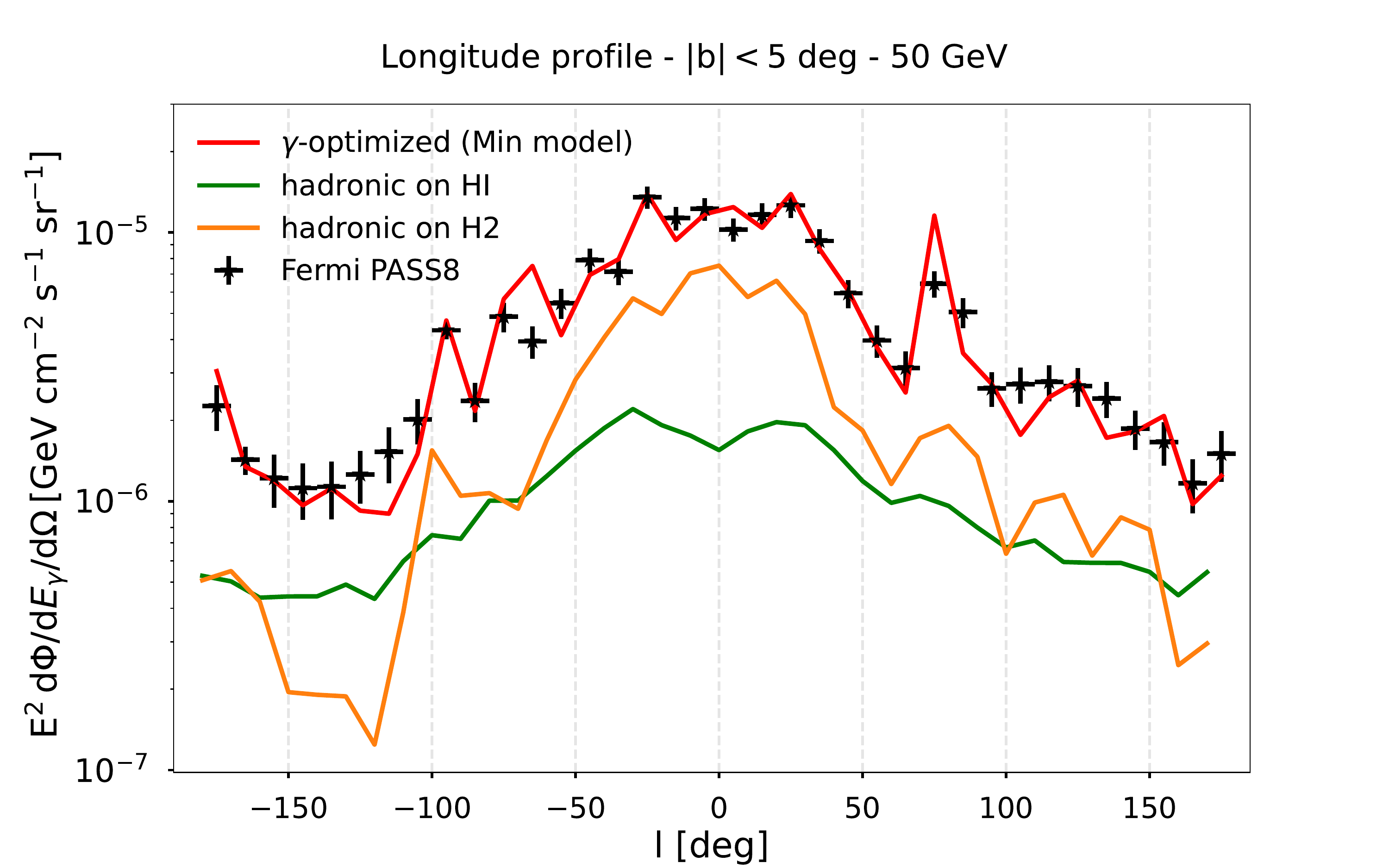}
\end{center}
\caption{Upper panels show the latitude profiles for the total $\gamma$-ray emission (for longitudes of $10^{\circ}$ $<$ $|l|$ $<$ $40^{\circ}$) at $100$ GeV predicted by the \textit{Base} and $\gamma$-\textit{optimized} models for the Max (left) and Min (right) configurations. 
Likewise, middle panels show longitude profiles at $100$ GeV predicted by the \textit{Base} and $\gamma$-\textit{optimized} models for the Max (left) and Min (right) configurations. 
The bottom panel reports the longitude $\gamma$-ray profile at $50$ GeV predicted by the $\gamma$-\textit{optimized} model (Min configuration), showing the emission that comes from the hadronic emission generated by HI (atomic) and H2 (molecular) gas. Fermi-LAT PASS8 data are added for comparison in every case.}\label{fig:Profile}
\end{figure}

Here we extend that analysis to other regions of the sky. 
First of all, in Figure~\ref{fig:Profile} we show latitudinal (top panels) and longitudinal (middle panels) profiles of the predicted $\gamma$-ray emission for the Min (right panels) and Max  (left panels) setups compared to \textit{Fermi}-LAT data.
%profiles mainly depend on the gas distribution model and demonstrate the good agreement between our model and the experimental data.
In this work, we make use of $\sim 149$ months of data (from 2008-08-04 to 2020-12-31), selecting CLEAN events from the PASS8 data.
Their extraction and calculation of exposure maps is performed using {\textit Fermi}-LAT ScienceTools\footnote{https://fermi.gsfc.nasa.gov/ssc/data/analysis/software/; https://github.com/fermi-lat/Fermitools-conda/wiki/Installation-Instructions} (2.0.8; released on 01/20/2021). We also account for the isotropic spectral template provided by the \textit{Fermi}-LAT collaboration~\cite{Fermi-LAT:2014ryh} (iso\_P8R3\_CLEAN\_V3\_v1)\footnote{https://fermi.gsfc.nasa.gov/ssc/data/access/lat/BackgroundModels.html}.

These plots show that the space-dependent $\gamma$-optimized scenario -- especially its Min setup which better traces the CR local spectra measured by CREAM and DAMPE -- provides a very good description of {\textit Fermi}-LAT results significantly better that the conventional Base one.
 In addition, in the bottom panel of the figure, we show the longitude profile of the emission at $50$~GeV specifying the hadronic emission originated from collisions of CRs with molecular gas (H2) and atomic gas (HI), in order to illustrate the importance of both contributions in different parts of the Galaxy. 

\begin{figure}[ht!]
\begin{center}
\includegraphics[width=0.65\linewidth]{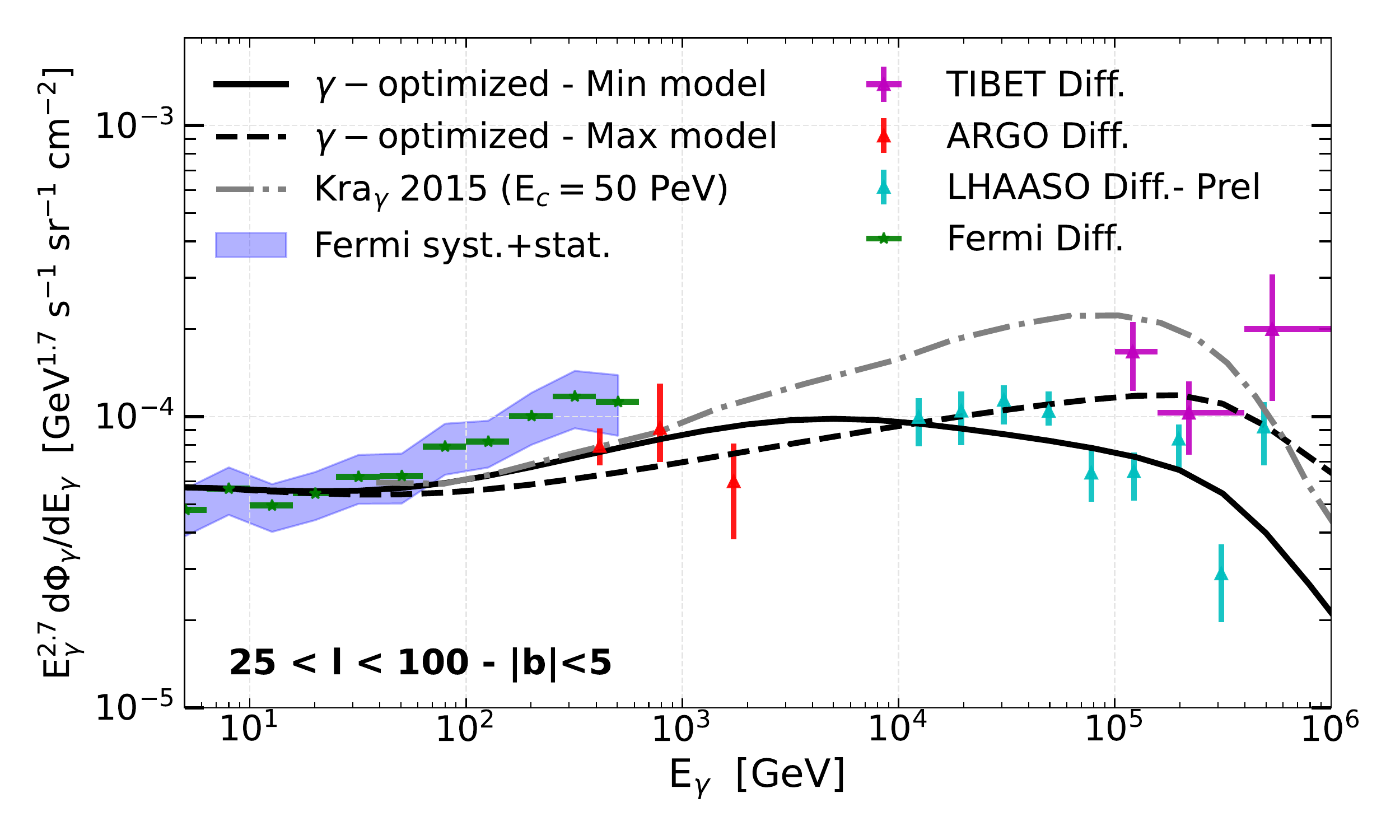}
\end{center}
\caption{ $\gamma$-ray diffuse spectra from the $\gamma$-optimized scenario compared to Tibet AS$\gamma$~\citep{TibetASgamma:2021tpz}, LHAASO~\citep{Zhao:2021dqj} (preliminary),  Fermi-LAT~\citep{Fermi-LAT:2012edv} (CLEAN events from PASS8 data with subtraction of flux from known sources and isotropic background)} and ARGO-YBJ~\citep{ARGO-YBJ:2015cpa} data in the window $| b | < 5^\circ $, $ 25^\circ < l < 100^\circ $. The KRA$_\gamma$ model (cutoff energy of E$_c=5$~PeV)~\citep{Gaggero:2015xza} is also included. Here, we account for absorption of $\gamma$-rays into CMB photons (see Fig.~7 of Ref.~\cite{Luque:2022buq}) and do not include the contribution of unresolved sources, which may be significant at the highest energies.
\label{fig:Gamma_Spectra}
\end{figure}

With this result at hand, we use the $\gamma$-optimized models to predict the $\gamma$-ray and $\nu$ spectra up to the PeV. 
For the $\gamma$-ray production cross-sections we used those by~\citet{Kelner2008prd} with the updated parameterization of the proton–proton total inelastic cross-section reported in~\citet{Kafexhiu2014prd}.
At those energies we need to account for $\gamma$-ray opacity due to the scattering onto CMB photons -- giving a $\sim 10\%$ depletion around the PeV -- having checked that the effect of the interstellar radiation fields is negligible. The comparison of the $\gamma$-ray flux from the $\gamma$-optimized scenario is shown in Figure~\ref{fig:Gamma_Spectra} in comparison to Tibet AS$\gamma$~\citep{TibetASgamma:2021tpz}, LHAASO~\citep{Zhao:2021dqj} (preliminary),  Fermi-LAT~\cite{Fermi-LAT:2012edv} and ARGO-YBJ~\citep{ARGO-YBJ:2015cpa} data in the window $| b | < 5^\circ $, $ 25^\circ < l < 100^\circ $. The KRA$_\gamma^5$ model (cutoff energy of E$_c=5$~PeV)~\citep{Gaggero:2015xza} is also included here for reference.

\section{Prospects for $\nu$ emission}

As anticipated above, one of the main consequences of the $\gamma$-\textit{optimized} model is that the hadronic $\gamma$-ray emission dominates over the leptonic one even at very high energies: in particular, it is expected to dominate in the innermost region of the GP. %and where over-densities of the interstellar medium are present.\\ 
While the observation of related very-high-energy $\gamma$-ray emission ($E \gg 10$ TeV) from this region is very challenging for the currently operating wide-field $\gamma$-ray observatories, mostly located in the Northern hemisphere, the $\nu$ emission can be observed from both hemispheres taking into account different event topologies and reconstruction strategies.

The diffuse Galactic $\nu$ emission is expected to overcome possible point-like excess in the inner GP region with spectral features inherited from the accelerated CR populations. With the $\nu$ data recorded in the last decade by IceCube and ANTARES, it was possible to constrain this important diffuse signal that can account for $\sim 10\%$ of the total astrophysical events collected by IceCube.

%This is especially relevant taking in mind that, differently from the running very high energy ($E \gg 10$ TeV) $\gamma$-ray observatories which are located in the Norther hemisphere,  $\nu$ telescopes in both hemispheres -- though with different reconstruction algorithms -- are expected to collect most of the diffuse emission from the GC region.  

Different studies from ANTARES and IceCube~\citep{ANTARES:2017nlh,ANTARES:2018nyb} placed upper limits on this diffuse $\nu$ emission taking advantage of the template fitting analysis method, with the model templates accurately reproducing the spatial distribution of the expected Galactic emission. This approach represented a step forward for the study of this diffuse signal considering that the angular resolution of different $\nu$ samples improved with time for both observatories thanks to new reconstruction quality techniques. Recently, an indication of a diffuse Galactic $\nu$ excess ($2 \sigma$) tracing the KRA$_\gamma^5$ template was reported by the IceCube collaboration \citep{IceCube:2019lzm} using seven years of collected cascade-type events.

%In fact, a ($2 \sigma$) hint was recently reported by the IceCube collaboration \citep{IceCube:2019lzm} in a combined analysis based on seven years of IceCube tracks
%and ten years of ANTARES tracks and showers using
%a likelihood ratio test. 
The main result of that calculation is a hint for a non-zero diffuse Galactic $\nu$ component, with a best-fit flux lying very close to the level of the $\gamma$-optimized model, with 29\% p-value (see Fig.~\ref{fig:Spectrum}). We remark that the $\gamma$-optimized model adopted in that work is based on \cite{Gaggero:2015xza}, which did not feature the improved modeling of the knee region presented in this work.

\begin{figure}[ht!]
\begin{center}
\includegraphics[width=8.5cm]{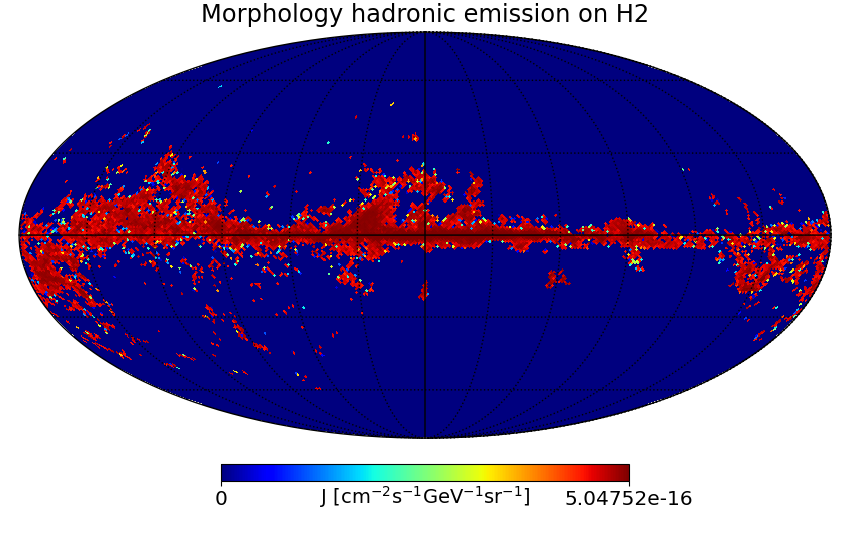}% 
\hspace{0.3cm}
\includegraphics[width=8.5cm]{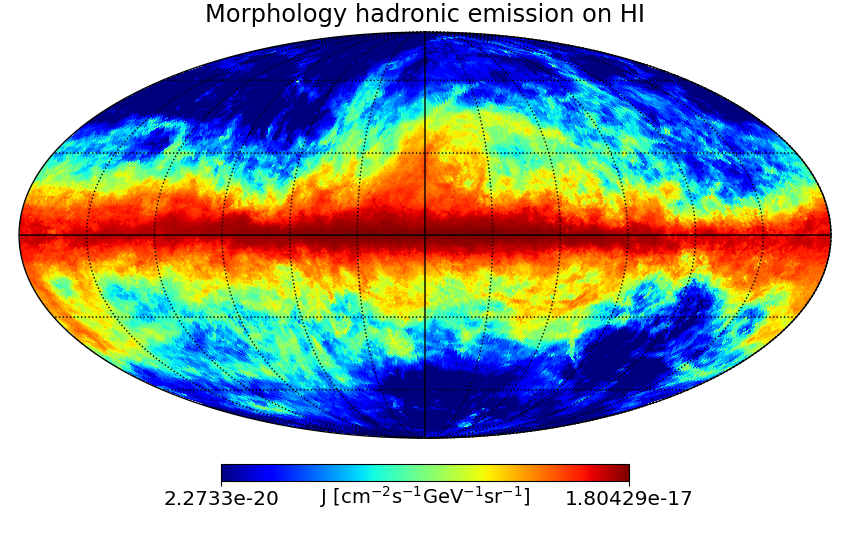}% 
\end{center}
\caption{ HEALPIX maps (NSIDE=512) showing the morphology of the hadronic emission for $100$~TeV $\gamma$-rays for the $\gamma$-\textit{optimized} model (Min configuration). The left map shows the hadronic emission generated by the interactions of CRs with molecular (H2) gas, while the right map shows the hadronic emission generated by interactions of CRs with atomic (HI) gas. This distribution will be also followed by the $\nu$ emission. }
\label{fig:map}
\end{figure}

\begin{figure}[ht!]
\begin{center}
\includegraphics[width=0.7\linewidth]{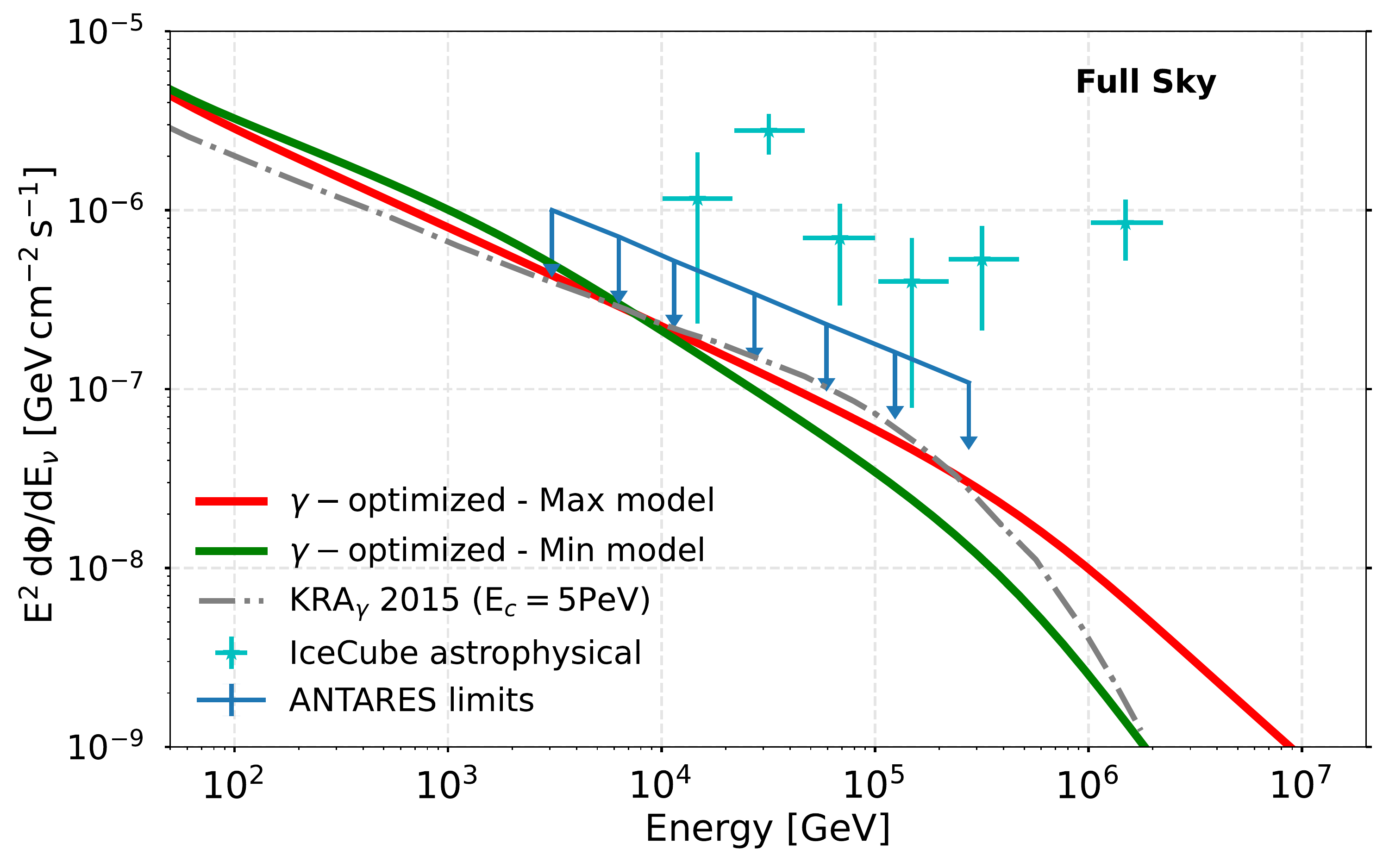}
\end{center}
\caption{ Full sky $\nu$ diffuse emission predicted from the $\gamma$-\textit{optimized} model (Min and Max configurations) compared to the model-independent upper limits obtained from the ANTARES collaboration. The predicted galactic $\nu$ flux from the KRA$_\gamma$ model (cutoff energy of E$_c=5$~PeV)~\citep{Gaggero:2015xza} is also included. In addition, the IceCube astrophysical $\nu$ flux as measured from IceCube using 7.5 years of track events~\citep{IceCube:2020wum} are added for completeness. }
\label{fig:Spectrum}
\end{figure}

In Figure~\ref{fig:map} we display the morphology of the hadronic emission for $100$~TeV $\gamma$-rays for the $\gamma$-\textit{optimized} model (Min configuration). This distribution will be also followed by the $\nu$ emission and it serves as a template to explore the zones where the $\nu$ emission will be more significant. As for the $\gamma$-ray production, we use the cross-sections described in~\cite{Kelner2008prd}, with inelastic cross-section from in~\citet{Kafexhiu2014prd}, for the $\nu$ emission. We have tested that other common parameterizations (namely, ~\cite{Kamae_2006} and AAFRAG~\cite{Koldobskiy:2021nld}) do not lead to important discrepancies at TeV energies, although a dedicated comparison of the effect of different cross sections in the predictions of the $\gamma$-ray and $\nu$ diffuse flux is left for a future work.

In Figure~\ref{fig:Spectrum} we show the predicted $\nu$ Galactic diffuse emission considering the Min and Max configurations of the $\gamma$-\textit{optimized} scenario and the expected Galactic $\nu$ flux from the KRA$_\gamma$ model (cutoff energy of E$_c=5$~PeV)~\citep{Gaggero:2015xza}, compared to the model-independent limits obtained from the ANTARES collaboration~\cite{ANTARES:2016mwq} considering six years of track-like events for the region $|l|<40^{\circ}$ and $|b|<3^{\circ}$. 
%We do not report the model-dependent upper limits based on KRA-$\gamma$ template \textbf{refs} since a new dedicated analysis will pose soon new constrains based on the novel $\gamma$-\textit{optimized} $\nu$ template.\\
We notice that our scenario is compatible with the current upper limits set by the ANTARES and IceCube collaborations, and that the Max model is particularly close to that limit. 
Moreover, since the spectral energy distributions of the KRA$_\gamma^5$ and $\gamma$-\textit{optimized} Max models are very close below 100 TeV, the hint of an excess tracing the former model found by the IceCube collaboration should hold also for the updated $\gamma$-\textit{optimized} Max model.  
%started to shed light on a Galactic signal that is now similarly estimated also by the $\gamma$-\textit{optimized} model. \\
%Moreover $\gamma$-\textit{optimized} model is compatible with recent observations the maximal energy reached by the Galactic CR accelerators seems to lie below tens of PeVs.
Remarkably, this is also the model which, for $\gamma$-rays, better matches the Tibet AS$\gamma$ results (see Fig.4 and 5 in~\citet{Luque:2022buq} and Fig.3 of~\citet{Eckner} ).

As also shown in Figure~\ref{fig:Spectrum} at energies larger than 10 TeV the $\gamma$-\textit{optimized} Min and Max configurations predict significantly different $\nu$ spectral shapes. Forthcoming IceCube or KM3NeT measurements should have the sensitivity to single out the correct between those models. We also notice that these results are compatible with the non-detection of galactic $\nu$ emission from Super-Kamiokande~\cite{Abe_2006}, since at the low energies at which Super-Kamiokande is able to perform the detection our model predicts a similar $\nu$ flux as the conventional models.

As already pointed out in \citet{Luque:2022buq} we notice that a degeneracy holds between the choice of the transport model and the shape of the source spectrum above 10 TeV. Indeed the spectrum predicted for Min configuration of the $\gamma$-\textit{optimized} scenario is quite close to the \textit{Base} Max model at those high energies. This degeneracy, however, can be broken using $\gamma$-ray data at lower energy showing, again, the importance of synergistically using $\nu$ and $\gamma$-ray channels to get to fully understanding of the underlying physics.

\section{Conclusions}

More than 100 years after the discovery of CRs, our current knowledge about the sources of these particles and the way they propagate through the Galaxy is still very limited. 
This is especially true at very high energies ($E \gg 100$ TeV) at which many interesting and enlightening  phenomena are expected to take place.   
This situation is swiftly improving thanks to the use of new $\gamma$-ray and $\nu$ measurements  which are providing complementary information besides that given by CR data alone. 
The consistent interpretation of those multi-messenger results requires accurate modeling of CR propagation and interaction with the ISM under more general conditions than those generally assumed to describe local CR data. 
%in combination with new CR data allow to better 
%local CR spectra allow us to improve our models of propagation of charged particles in the Galaxy and solve this puzzle.

In this paper, we have revisited a recently released CR transport model implemented to allow a consistent description of local CR data up to energies of several PeVs and the diffuse emission $\gamma$-ray spectrum measured by {\textit Fermi}-LAT, ARGO and Tibet AS$\gamma$ in different regions of the Galaxy.

%in different regions of the galactic plane with the local measurements of CR fluxes is exploited in order to constrain the spatial dependence of the diffusion coefficient that governs the transport of these particles. 
In particular, we have reported new details on the predicted CR spectra in different regions of the Galactic plane obtained from the $\gamma$-\textit{optimized} (spatial dependent transport) scenario and shown that they are totally compatible with the local measurements of the boron-over-carbon ratio (B/C) from the GeV region.
In addition, we have compared the predicted $\gamma$-ray diffuse emission latitude and longitude profiles with the \textit{Fermi}-LAT PASS8 data. We also showed the contributions of the hadronic emission coming from the different phases of the gas, namely atomic and molecular gas. A good agreement is found throughout the full plane of the Galaxy thanks to the CR density enhancement predicted by this scenario in the innermost region of the Galactic plane around 100 GeV. 

At larger energies the predictions of our scenario depend on the assumed source spectral shape. Due to the large experimental uncertainties above 10 TeV we considered two configurations (Min and Max) roughly bracketing the available proton and Helium experimental data.  

For those models we computed here for the first time -- though similarly to what already done for $\gamma$-rays -- the diffuse $\nu$ emission of the Galaxy over a wide energy range which we compared with available ANTARES and IceCube upper limits and provide as templates for the forthcoming experimental campaigns. 

We showed that the predicted full-sky spectrum for the Max configuration is very close to that obtained with the KRA$_\gamma^5$ model presented in \citet{Gaggero:2015xza}.
Since this model has recently received a positive -- though not yet conclusive -- evidence
by the likelyhood analysis of the cascade events collected by IceCube \citep{IceCube:2019lzm} our results provide a very intriguing science case for future analyses, which may have the unique opportunity to confirm this prediction.

We argued that complementary analysis of $\gamma$-ray add $\nu$ emissions are required both 
to lift the degeneracy between the choice of the transport scenario and the shape of the source spectra at very high energies and to constrain the contribution of unresolved sources to the $\gamma$-ray diffuse emission of the Galaxy.

%Our study lead to full-sky maps of the truly diffuse $\gamma$-ray and $\nu$ emissions over a wide energy range that we make publicly available and can be used as high resolution templates for new generation of high energy $\gamma$-ray and $\nu$ observatories.

%and, respect to conventional transport models,  predicts an important enhancement of the hadronic interactions towards the Galactic Centre.

%Therefore, we reported high resolution maps that show the morphology of the hadronic emission from the $\gamma$-\textit{optimized} model (Min configuration), separately for the atomic and molecular gas at 100 TeV. The angular distribution of the diffuse $\nu$ emission traces that of the $\gamma$-ray hadronic emission and these maps can be used as templates for likelihood analysis by experimental collaborations in any selected region of the sky. 

%A testable signature of this model will be the $\nu$ emission, that also will help us discern better between a leptonic (unresolved sources) and hadronic (the truly diffuse) origin of the $\gamma$-ray emission in different parts of the Galaxy. 

We, finally, show here the all-sky predicted $\nu$ emission from this model, compared to upper limits from the ANTARES collaboration, showing that a possible measurement of this emission can be really around the corner. We also emphasize that the different observations coming from high-energy CRs, $\gamma$-rays and $\nu$ seem to be compatible with a maximal energy reached by the Galactic CR accelerators below tens of PeV. The diffuse $\nu$ emission throughout the Galaxy and its possible experimental detection will be extensively explored in a follow-up paper.

\newpage

\section*{Conflict of Interest Statement}
The authors declare that the research was conducted in the absence of any commercial or financial relationships that could be construed as a potential conflict of interest.

\section*{Author Contributions}

Pedro de la Torre has been in charge of the technical calculations. Daniele Gaggero, Dario Grasso and Antonio Marinelli has been critically revising the ideas involved and discussing ways to improve the presentation of the materials shown here.
All the authors have made a substantial contribution to the concept and design of the article and drafted the article with important intellectual content.

\section*{Funding}
P. De la Torre is supported by the Swedish National Space Agency under contract 117/19.

D. Gaggero acknowledges support from Generalitat Valenciana through the plan GenT program (CIDEGENT/2021/017).
C.E.~acknowledges the European Commission for support under the H2020-MSCA-IF-2016 action, Grant No.~751311 GRAPES 8211 Galactic cosmic RAy Propagation: An Extensive Study.

\section*{Acknowledgments}
This work couldn't be carried out without the help of Ottavio fornieri, Carmelo Evoli, Kathrin Egberts and Constantin Steppa. We thank Quentin Remy for providing us with the interstellar HI and H$_2$ 3D distributions (``ring model'') used in this work. We also thank Hershal Pandya for informing us about IceCube collaboration $\gamma$-ray measurements and providing us with the corresponding sky window. 
We thank Paolo Lipari and Silvia Vernetto as well, for reading our manuscript and giving us useful comments.
This project used computing resources from the Swedish National Infrastructure for Computing (SNIC) under project Nos. 2021/3-42, 2021/6-326 and 2021-1-24 partially funded by the Swedish Research Council through grant no. 2018-05973.

%\section*{Supplemental Data}
% \href{http://home.frontiersin.org/about/author-guidelines#SupplementaryMaterial}{Supplementary Material} should be uploaded separately on submission, if there are Supplementary Figures, please include the caption in the same file as the figure. LaTeX Supplementary Material templates can be found in the Frontiers LaTeX folder.

\section*{Data Availability Statement}
The datasets analyzed to generate our models can be found at \url{https://heasarc.gsfc.nasa.gov/FTP/fermi/data/lat/weekly/}.
The diffuse $\gamma$-ray maps produced in this paper can be found and downloaded in the repository \url{https://github.com/tospines/Gamma-variable_High-resolution}

\bibliographystyle{Frontiers-Harvard} 
\bibliography{test}

\begin{thebibliography}{48}
\providecommand{\natexlab}[1]{#1}
\expandafter\ifx\csname urlstyle\endcsname\relax
  \providecommand{\doi}[1]{doi:\discretionary{}{}{}#1}\else
  \providecommand{\doi}{doi:\discretionary{}{}{}\begingroup
  \urlstyle{rm}\Url}\fi
\providecommand{\selectlanguage}[1]{\relax}
\providecommand{\bibAnnoteFile}[1]{%
  \IfFileExists{#1}{\begin{quotation}\noindent\textsc{Key:} #1\\
  \textsc{Annotation:}\ \input{#1}\end{quotation}}{}}
\providecommand{\bibAnnote}[2]{%
  \begin{quotation}\noindent\textsc{Key:} #1\\
  \textsc{Annotation:}\ #2\end{quotation}}

\bibitem[{Aartsen et~al.(2014)}]{IceCube:2014stg}
Aartsen, M.~G. et~al. (2014).
\newblock {Observation of High-Energy Astrophysical Neutrinos in Three Years of
  IceCube Data}.
\newblock \emph{Phys. Rev. Lett.} 113, 101101.
\newblock \doi{10.1103/PhysRevLett.113.101101}
\bibAnnoteFile{IceCube:2014stg}

\bibitem[{Aartsen et~al.(2019)}]{IceCube:2019lzm}
Aartsen, M.~G. et~al. (2019).
\newblock {Search for Sources of Astrophysical Neutrinos Using Seven Years of
  IceCube Cascade Events}.
\newblock \emph{Astrophys. J.} 886, 12.
\newblock \doi{10.3847/1538-4357/ab4ae2}
\bibAnnoteFile{IceCube:2019lzm}

\bibitem[{Abbasi et~al.(2021)}]{IceCube:2020wum}
Abbasi, R. et~al. (2021).
\newblock {The IceCube high-energy starting event sample: Description and flux
  characterization with 7.5 years of data}.
\newblock \emph{Phys. Rev. D} 104, 022002.
\newblock \doi{10.1103/PhysRevD.104.022002}
\bibAnnoteFile{IceCube:2020wum}

\bibitem[{Abe et~al.(2006)Abe, Hosaka, Iida, Ishihara, Kameda, Koshio
  et~al.}]{Abe_2006}
Abe, K., Hosaka, J., Iida, T., Ishihara, K., Kameda, J., Koshio, Y., et~al.
  (2006).
\newblock High-energy neutrino astronomy using upward-going muons in
  super-kamiokande i.
\newblock \emph{The Astrophysical Journal} 652, 198--205.
\newblock \doi{10.1086/508016}
\bibAnnoteFile{Abe_2006}

\bibitem[{{Acero} et~al.(2016){Acero}, {Ackermann}, {Ajello}, {Albert},
  {Baldini}, {Ballet} et~al.}]{Acero2016apjs}
{Acero}, F., {Ackermann}, M., {Ajello}, M., {Albert}, A., {Baldini}, L.,
  {Ballet}, J., et~al. (2016).
\newblock {Development of the Model of Galactic Interstellar Emission for
  Standard Point-source Analysis of Fermi Large Area Telescope Data}.
\newblock \emph{apjs} 223, 26.
\newblock \doi{10.3847/0067-0049/223/2/26}
\bibAnnoteFile{Acero2016apjs}

\bibitem[{Ackermann et~al.(2012)}]{Fermi-LAT:2012edv}
Ackermann, M. et~al. (2012).
\newblock {Fermi-LAT Observations of the Diffuse Gamma-Ray Emission:
  Implications for Cosmic Rays and the Interstellar Medium}.
\newblock \emph{Astrophys. J.} 750, 3.
\newblock \doi{10.1088/0004-637X/750/1/3}
\bibAnnoteFile{Fermi-LAT:2012edv}

\bibitem[{Ackermann et~al.(2015)}]{Fermi-LAT:2014ryh}
Ackermann, M. et~al. (2015).
\newblock {The spectrum of isotropic diffuse gamma-ray emission between 100 MeV
  and 820 GeV}.
\newblock \emph{Astrophys. J.} 799, 86.
\newblock \doi{10.1088/0004-637X/799/1/86}
\bibAnnoteFile{Fermi-LAT:2014ryh}

\bibitem[{{Adri{\'a}n-Mart{\'\i}nez} et~al.(2016){Adri{\'a}n-Mart{\'\i}nez},
  {Ageron}, {Aharonian}, {Aiello}, {Albert}, {Ameli} et~al.}]{KM3NET2016}
{Adri{\'a}n-Mart{\'\i}nez}, S., {Ageron}, M., {Aharonian}, F., {Aiello}, S.,
  {Albert}, A., {Ameli}, F., et~al. (2016).
\newblock {Letter of intent for KM3NeT 2.0}.
\newblock \emph{Journal of Physics G Nuclear Physics} 43, 084001.
\newblock \doi{10.1088/0954-3899/43/8/084001}
\bibAnnoteFile{KM3NET2016}

\bibitem[{Adrian-Martinez et~al.(2016)}]{ANTARES:2016mwq}
Adrian-Martinez, S. et~al. (2016).
\newblock {Constraints on the neutrino emission from the Galactic Ridge with
  the ANTARES telescope}.
\newblock \emph{Phys. Lett. B} 760, 143--148.
\newblock \doi{10.1016/j.physletb.2016.06.051}
\bibAnnoteFile{ANTARES:2016mwq}

\bibitem[{Adriani et~al.(2011)Adriani, Barbarino, Bazilevskaya, Bellotti,
  Boezio, Bogomolov et~al.}]{doi:10.1126/science.1199172}
Adriani, O., Barbarino, G.~C., Bazilevskaya, G.~A., Bellotti, R., Boezio, M.,
  Bogomolov, E.~A., et~al. (2011).
\newblock Pamela measurements of cosmic-ray proton and helium spectra.
\newblock \emph{Science} 332, 69--72.
\newblock \doi{10.1126/science.1199172}
\bibAnnoteFile{doi:10.1126/science.1199172}

\bibitem[{Aguilar et~al.(2015)Aguilar, Aisa, Alpat, Alvino, Ambrosi, Andeen
  et~al.}]{PhysRevLett.114.171103}
Aguilar, M., Aisa, D., Alpat, B., Alvino, A., Ambrosi, G., Andeen, K., et~al.
  (2015).
\newblock Precision measurement of the proton flux in primary cosmic rays from
  rigidity 1 gv to 1.8 tv with the alpha magnetic spectrometer on the
  international space station.
\newblock \emph{Phys. Rev. Lett.} 114, 171103.
\newblock \doi{10.1103/PhysRevLett.114.171103}
\bibAnnoteFile{PhysRevLett.114.171103}

\bibitem[{Ahlers et~al.(2016)Ahlers, Bai, Barger, and Lu}]{Ahlers:2015moa}
Ahlers, M., Bai, Y., Barger, V., and Lu, R. (2016).
\newblock {Galactic neutrinos in the TeV to PeV range}.
\newblock \emph{Phys. Rev. D} 93, 013009.
\newblock \doi{10.1103/PhysRevD.93.013009}
\bibAnnoteFile{Ahlers:2015moa}

\bibitem[{Ahn et~al.(2010)Ahn, Allison, Bagliesi, Beatty, Bigongiari, Childers
  et~al.}]{Ahn_2010}
Ahn, H.~S., Allison, P., Bagliesi, M.~G., Beatty, J.~J., Bigongiari, G.,
  Childers, J.~T., et~al. (2010).
\newblock {DISCREPANT} {HARDENING} {OBSERVED} {IN} {COSMIC}-{RAY} {ELEMENTAL}
  {SPECTRA}.
\newblock \emph{The Astrophysical Journal} 714, L89--L93.
\newblock \doi{10.1088/2041-8205/714/1/l89}
\bibAnnoteFile{Ahn_2010}

\bibitem[{Albert et~al.(2017)}]{ANTARES:2017nlh}
Albert, A. et~al. (2017).
\newblock {New constraints on all flavor Galactic diffuse neutrino emission
  with the ANTARES telescope}.
\newblock \emph{Phys. Rev. D} 96, 062001.
\newblock \doi{10.1103/PhysRevD.96.062001}
\bibAnnoteFile{ANTARES:2017nlh}

\bibitem[{Albert et~al.(2018)}]{ANTARES:2018nyb}
Albert, A. et~al. (2018).
\newblock {Joint Constraints on Galactic Diffuse Neutrino Emission from the
  ANTARES and IceCube Neutrino Telescopes}.
\newblock \emph{Astrophys. J. Lett.} 868, L20.
\newblock \doi{10.3847/2041-8213/aaeecf}
\bibAnnoteFile{ANTARES:2018nyb}

\bibitem[{Amenomori et~al.(2021)}]{TibetASgamma:2021tpz}
Amenomori, M. et~al. (2021).
\newblock {First Detection of sub-PeV Diffuse Gamma Rays from the Galactic
  Disk: Evidence for Ubiquitous Galactic Cosmic Rays beyond PeV Energies}.
\newblock \emph{Phys. Rev. Lett.} 126, 141101.
\newblock \doi{10.1103/PhysRevLett.126.141101}
\bibAnnoteFile{TibetASgamma:2021tpz}

\bibitem[{Bartoli et~al.(2015)}]{ARGO-YBJ:2015cpa}
Bartoli, B. et~al. (2015).
\newblock {Study of the Diffuse Gamma-ray Emission From the Galactic Plane With
  ARGO-YBJ}.
\newblock \emph{Astrophys. J.} 806, 20.
\newblock \doi{10.1088/0004-637X/806/1/20}
\bibAnnoteFile{ARGO-YBJ:2015cpa}

\bibitem[{Berezinsky et~al.(1993)Berezinsky, Gaisser, Halzen, and
  Stanev}]{Berezinsky:1992wr}
Berezinsky, V.~S., Gaisser, T.~K., Halzen, F., and Stanev, T. (1993).
\newblock {Diffuse radiation from cosmic ray interactions in the galaxy}.
\newblock \emph{Astropart. Phys.} 1, 281--288.
\newblock \doi{10.1016/0927-6505(93)90014-5}
\bibAnnoteFile{Berezinsky:1992wr}

\bibitem[{Blasi et~al.(2012)Blasi, Amato, and Serpico}]{Blasi2012}
Blasi, P., Amato, E., and Serpico, P.~D. (2012).
\newblock Spectral breaks as a signature of cosmic ray induced turbulence in
  the galaxy.
\newblock \emph{Phys. Rev. Lett.} 109, 061101.
\newblock \doi{10.1103/PhysRevLett.109.061101}
\bibAnnoteFile{Blasi2012}

\bibitem[{Casanova and Dingus(2008)}]{Casanova:2007cf}
Casanova, S. and Dingus, B.~L. (2008).
\newblock {Constraints on the TeV source population and its contribution to the
  galactic diffuse TeV emission}.
\newblock \emph{Astropart. Phys.} 29, 63--69.
\newblock \doi{10.1016/j.astropartphys.2007.11.008}
\bibAnnoteFile{Casanova:2007cf}

\bibitem[{{Cerri} et~al.(2017){Cerri}, {Gaggero}, {Vittino}, {Evoli}, and
  {Grasso}}]{Cerri2017jcap}
{Cerri}, S.~S., {Gaggero}, D., {Vittino}, A., {Evoli}, C., and {Grasso}, D.
  (2017).
\newblock {A signature of anisotropic cosmic-ray transport in the gamma-ray
  sky}.
\newblock \emph{jcap} 2017, 019.
\newblock \doi{10.1088/1475-7516/2017/10/019}
\bibAnnoteFile{Cerri2017jcap}

\bibitem[{{Cesarsky} and {Montmerle}(1983)}]{1983SSRv...36..173C}
{Cesarsky}, C.~J. and {Montmerle}, T. (1983).
\newblock {Gamma-Rays from Active Regions in the Galaxy - the Possible
  Contribution of Stellar Winds}.
\newblock \emph{ssr} 36, 173--193.
\newblock \doi{10.1007/BF00167503}
\bibAnnoteFile{1983SSRv...36..173C}

\bibitem[{de~la Torre~Luque et~al.(2022)de~la Torre~Luque, Mazziotta, Ferrari,
  Loparco, Sala, and Serini}]{delaTorreLuque:2022vhm}
de~la Torre~Luque, P., Mazziotta, M.~N., Ferrari, A., Loparco, F., Sala, P.,
  and Serini, D. (2022).
\newblock {FLUKA cross sections for cosmic-ray interactions with the DRAGON2
  code}.
\newblock \emph{JCAP} 07, 008.
\newblock \doi{10.1088/1475-7516/2022/07/008}
\bibAnnoteFile{delaTorreLuque:2022vhm}

\bibitem[{Dundovic et~al.(2021)Dundovic, Evoli, Gaggero, and
  Grasso}]{Dundovic:2021ryb}
Dundovic, A., Evoli, C., Gaggero, D., and Grasso, D. (2021).
\newblock {Simulating the Galactic multi-messenger emissions with HERMES}.
\newblock \emph{Astron. Astrophys.} 653, A18.
\newblock \doi{10.1051/0004-6361/202140801}
\bibAnnoteFile{Dundovic:2021ryb}

\bibitem[{Eckner and Calore(2022)}]{Eckner}
[Dataset] Eckner, C. and Calore, F. (2022).
\newblock First constraints on axion-like particles from galactic sub-pev gamma
  rays.
\newblock \doi{10.48550/ARXIV.2204.12487}
\bibAnnoteFile{Eckner}

\bibitem[{{Evoli} et~al.(2017){Evoli}, {Gaggero}, {Vittino}, {Di Bernardo}, {Di
  Mauro}, {Ligorini} et~al.}]{Evoli2017jcap}
{Evoli}, C., {Gaggero}, D., {Vittino}, A., {Di Bernardo}, G., {Di Mauro}, M.,
  {Ligorini}, A., et~al. (2017).
\newblock {Cosmic-ray propagation with DRAGON2: I. numerical solver and
  astrophysical ingredients}.
\newblock \emph{jcap} 2017, 015.
\newblock \doi{10.1088/1475-7516/2017/02/015}
\bibAnnoteFile{Evoli2017jcap}

\bibitem[{{Evoli} et~al.(2018){Evoli}, {Gaggero}, {Vittino}, {Di Mauro},
  {Grasso}, and {Mazziotta}}]{Evoli2018jcap}
{Evoli}, C., {Gaggero}, D., {Vittino}, A., {Di Mauro}, M., {Grasso}, D., and
  {Mazziotta}, M.~N. (2018).
\newblock {Cosmic-ray propagation with DRAGON2: II. Nuclear interactions with
  the interstellar gas}.
\newblock \emph{jcap} 2018, 006.
\newblock \doi{10.1088/1475-7516/2018/07/006}
\bibAnnoteFile{Evoli2018jcap}

\bibitem[{Evoli et~al.(2007)Evoli, Grasso, and Maccione}]{Evoli:2007iy}
Evoli, C., Grasso, D., and Maccione, L. (2007).
\newblock {Diffuse Neutrino and Gamma-ray Emissions of the Galaxy above the
  TeV}.
\newblock \emph{JCAP} 06, 003.
\newblock \doi{10.1088/1475-7516/2007/06/003}
\bibAnnoteFile{Evoli:2007iy}

\bibitem[{Gaggero et~al.(2015{\natexlab{a}})Gaggero, Grasso, Marinelli, Urbano,
  and Valli}]{Gaggero:2015xza}
Gaggero, D., Grasso, D., Marinelli, A., Urbano, A., and Valli, M.
  (2015{\natexlab{a}}).
\newblock {The gamma-ray and neutrino sky: A consistent picture of Fermi-LAT,
  Milagro, and IceCube results}.
\newblock \emph{Astrophys. J. Lett.} 815, L25.
\newblock \doi{10.1088/2041-8205/815/2/L25}
\bibAnnoteFile{Gaggero:2015xza}

\bibitem[{Gaggero et~al.(2015{\natexlab{b}})Gaggero, Urbano, Valli, and
  Ullio}]{Gaggero:2014xla}
Gaggero, D., Urbano, A., Valli, M., and Ullio, P. (2015{\natexlab{b}}).
\newblock {Gamma-ray sky points to radial gradients in cosmic-ray transport}.
\newblock \emph{Phys. Rev. D} 91, 083012.
\newblock \doi{10.1103/PhysRevD.91.083012}
\bibAnnoteFile{Gaggero:2014xla}

\bibitem[{Guo and Yuan(2018)}]{GuoModel}
Guo, Y.-Q. and Yuan, Q. (2018).
\newblock Understanding the spectral hardenings and radial distribution of
  galactic cosmic rays and fermi diffuse $\ensuremath{\gamma}$ rays with
  spatially-dependent propagation.
\newblock \emph{Phys. Rev. D} 97, 063008.
\newblock \doi{10.1103/PhysRevD.97.063008}
\bibAnnoteFile{GuoModel}

\bibitem[{{Kafexhiu} et~al.(2014){Kafexhiu}, {Aharonian}, {Taylor}, and
  {Vila}}]{Kafexhiu2014prd}
{Kafexhiu}, E., {Aharonian}, F., {Taylor}, A.~M., and {Vila}, G.~S. (2014).
\newblock {Parametrization of gamma-ray production cross sections for p p
  interactions in a broad proton energy range from the kinematic threshold to
  PeV energies}.
\newblock \emph{prd} 90, 123014.
\newblock \doi{10.1103/PhysRevD.90.123014}
\bibAnnoteFile{Kafexhiu2014prd}

\bibitem[{{Kamae} et~al.(2006){Kamae}, {Karlsson}, {Mizuno}, {Abe}, and
  {Koi}}]{Kamae_2006}
{Kamae}, T., {Karlsson}, N., {Mizuno}, T., {Abe}, T., and {Koi}, T. (2006).
\newblock {Parameterization of {\ensuremath{\gamma}}, e$^{+/-}$, and Neutrino
  Spectra Produced by p-p Interaction in Astronomical Environments}.
\newblock \emph{apj} 647, 692--708.
\newblock \doi{10.1086/505189}
\bibAnnoteFile{Kamae_2006}

\bibitem[{{Kelner} and {Aharonian}(2008)}]{Kelner2008prd}
{Kelner}, S.~R. and {Aharonian}, F.~A. (2008).
\newblock {Energy spectra of gamma rays, electrons, and neutrinos produced at
  interactions of relativistic protons with low energy radiation}.
\newblock \emph{prd} 78, 034013.
\newblock \doi{10.1103/PhysRevD.78.034013}
\bibAnnoteFile{Kelner2008prd}

\bibitem[{Kissmann(2014)}]{KISSMANN201437}
Kissmann, R. (2014).
\newblock Picard: A novel code for the galactic cosmic ray propagation problem.
\newblock \emph{Astroparticle Physics} 55, 37--50.
\newblock \doi{https://doi.org/10.1016/j.astropartphys.2014.02.002}
\bibAnnoteFile{KISSMANN201437}

\bibitem[{Koldobskiy et~al.(2021)Koldobskiy, Kachelrie\ss{}, Lskavyan, Neronov,
  Ostapchenko, and Semikoz}]{Koldobskiy:2021nld}
Koldobskiy, S., Kachelrie\ss{}, M., Lskavyan, A., Neronov, A., Ostapchenko, S.,
  and Semikoz, D.~V. (2021).
\newblock {Energy spectra of secondaries in proton-proton interactions}.
\newblock \emph{Phys. Rev. D} 104, 123027.
\newblock \doi{10.1103/PhysRevD.104.123027}
\bibAnnoteFile{Koldobskiy:2021nld}

\bibitem[{{Lagage} and {Cesarsky}(1983)}]{1983A&A...125..249L}
{Lagage}, P.~O. and {Cesarsky}, C.~J. (1983).
\newblock {The maximum energy of cosmic rays accelerated by supernova shocks.}
\newblock \emph{aap} 125, 249--257
\bibAnnoteFile{1983A&A...125..249L}

\bibitem[{Lipari and Vernetto(2018)}]{Lipari:2018gzn}
Lipari, P. and Vernetto, S. (2018).
\newblock {Diffuse Galactic gamma ray flux at very high energy}.
\newblock \emph{Phys. Rev. D} 98, 043003.
\newblock \doi{10.1103/PhysRevD.98.043003}
\bibAnnoteFile{Lipari:2018gzn}

\bibitem[{Luque(2021)}]{Luque:2021ddh}
Luque, P. D. L.~T. (2021).
\newblock {Combined analyses of the antiproton production from cosmic-ray
  interactions and its possible dark matter origin}.
\newblock \emph{JCAP} 11, 018.
\newblock \doi{10.1088/1475-7516/2021/11/018}
\bibAnnoteFile{Luque:2021ddh}

\bibitem[{Luque et~al.(2022)Luque, Gaggero, Grasso, Fornieri, Egberts, Steppa
  et~al.}]{Luque:2022buq}
Luque, P. D. l.~T., Gaggero, D., Grasso, D., Fornieri, O., Egberts, K., Steppa,
  C., et~al. (2022).
\newblock {Galactic diffuse gamma rays meet the PeV frontier}.
\newblock \emph{ArXiv:2203.15759}
\bibAnnoteFile{Luque:2022buq}

\bibitem[{Luque et~al.(2021)Luque, Mazziotta, Loparco, Gargano, and
  Serini}]{Luque:2021nxb}
Luque, P. D. L.~T., Mazziotta, M.~N., Loparco, F., Gargano, F., and Serini, D.
  (2021).
\newblock {Markov chain Monte Carlo analyses of the flux ratios of B, Be and Li
  with the DRAGON2 code}.
\newblock \emph{JCAP} 07, 010.
\newblock \doi{10.1088/1475-7516/2021/07/010}
\bibAnnoteFile{Luque:2021nxb}

\bibitem[{null null et~al.(2019)null null, An, Asfandiyarov, Azzarello,
  Bernardini, Bi et~al.}]{DAMPEBUMP}
null null, An, Q., Asfandiyarov, R., Azzarello, P., Bernardini, P., Bi, X.~J.,
  et~al. (2019).
\newblock Measurement of the cosmic ray proton spectrum from 40 gev to 100 tev
  with the dampe satellite.
\newblock \emph{Science Advances} 5, eaax3793.
\newblock \doi{10.1126/sciadv.aax3793}
\bibAnnoteFile{DAMPEBUMP}

\bibitem[{Strong and Moskalenko(1998)}]{Strong_1998}
Strong, A.~W. and Moskalenko, I.~V. (1998).
\newblock Propagation of cosmic-ray nucleons in the galaxy.
\newblock \emph{The Astrophysical Journal} 509, 212--228.
\newblock \doi{10.1086/306470}
\bibAnnoteFile{Strong_1998}

\bibitem[{{Thoudam, S.} et~al.(2016){Thoudam, S.}, {Rachen, J. P.}, {van Vliet,
  A.}, {Achterberg, A.}, {Buitink, S.}, {Falcke, H.} et~al.}]{Knee}
{Thoudam, S.}, {Rachen, J. P.}, {van Vliet, A.}, {Achterberg, A.}, {Buitink,
  S.}, {Falcke, H.}, et~al. (2016).
\newblock Cosmic-ray energy spectrum and composition up to the ankle: the case
  for a second galactic component.
\newblock \emph{A\&A} 595, A33.
\newblock \doi{10.1051/0004-6361/201628894}
\bibAnnoteFile{Knee}

\bibitem[{Vecchiotti et~al.(2021)Vecchiotti, Zuccarini, Villante, and
  Pagliaroli}]{Vecchiotti:2021yzk}
Vecchiotti, V., Zuccarini, F., Villante, F.~L., and Pagliaroli, G. (2021).
\newblock {Unresolved sources naturally contribute to PeV $\gamma$-ray diffuse
  emission observed by Tibet AS$\gamma$}
\bibAnnoteFile{Vecchiotti:2021yzk}

\bibitem[{{Yang} et~al.(2016){Yang}, {Aharonian}, and {Evoli}}]{Yang2016prd}
{Yang}, R., {Aharonian}, F., and {Evoli}, C. (2016).
\newblock {Radial distribution of the diffuse {\ensuremath{\gamma}}-ray
  emissivity in the Galactic disk}.
\newblock \emph{prd} 93, 123007.
\newblock \doi{10.1103/PhysRevD.93.123007}
\bibAnnoteFile{Yang2016prd}

\bibitem[{Yoon et~al.(2017)Yoon, Anderson, Barrau, Conklin, Coutu, Derome
  et~al.}]{Yoon_2017}
Yoon, Y.~S., Anderson, T., Barrau, A., Conklin, N.~B., Coutu, S., Derome, L.,
  et~al. (2017).
\newblock Proton and helium spectra from the {CREAM}-{III} flight.
\newblock \emph{The Astrophysical Journal} 839, 5.
\newblock \doi{10.3847/1538-4357/aa68e4}
\bibAnnoteFile{Yoon_2017}

\bibitem[{Zhao et~al.(2021)Zhao, Zhang, Zhang, and Yuan}]{Zhao:2021dqj}
Zhao, S., Zhang, R., Zhang, Y., and Yuan, Q. (2021).
\newblock {Measurement of the diffuse gamma-ray emission from Galactic plane
  with LHAASO-KM2A}.
\newblock \emph{PoS} ICRC2021, 859.
\newblock \doi{10.22323/1.395.0859}
\bibAnnoteFile{Zhao:2021dqj}

\end{thebibliography}

%\section*{Figures}

%%% Please be aware that for original research articles we only permit a combined number of 15 figures and tables, one figure with multiple subfigures will count as only one figure.

\textbf{Morphology of the hadronic emission. }

%%% If you are submitting a figure with subfigures please combine these into one image file with part labels integrated.
%%% If you don't add the figures in the LaTeX files, please upload them when submitting the article.
%%% Frontiers will add the figures at the end of the provisional pdf automatically
%%% The use of LaTeX coding to draw Diagrams/Figures/Structures should be avoided. They should be external callouts including graphics.

\end{document}